\RequirePackage[2020-02-02]{latexrelease}
\documentclass[%
 reprint,
 amsmath,amssymb,
 aps,
]{revtex4-2}
\usepackage{mathrsfs}
\usepackage{longtable}
\usepackage{graphicx}
\usepackage{dcolumn}
\usepackage{bm}
\usepackage{multirow}

\begin{document}
	\preprint{APS/123-QED}
	
%
		\title{Relativistic Flux-Tube Model Predictions from Charmed Mesons to Double-Charmed Baryons} 
%
	\author{Pooja Jakhad$ ^{1*} $}
	\author{Ajay Kumar Rai$ ^{1} $}
	\affiliation{$ ^{1*} $Department of Physics, Sardar Vallabhbhai National Institute of Technology, Surat, Gujarat-395007, India}
	
	
	\date{\today}
	
\begin{abstract}
Utilizing comprehensive experimental data on charmed mesons, we systematically investigate masses of the higher radial and orbital excitations of the $D$ and $D_s$ meson families using the relativistic flux tube model. Our study employs mass splitting induced by spin-dependent interactions within the j–j coupling scheme. Our predicted masses align well with the experimental measurements for the well-established $D$ and $D_s$ states. However, anomalous resonances such as $D_{s0}(2317)$ and $D_{s1}(2460)$ do not align with conventional meson states within our theoretical framework. 
{We also study strong decays using an effective Lagrangian formalism that combines heavy-quark symmetry with chiral dynamics, focusing on two-body modes into a charmed meson plus a light pseudoscalar ($\pi$, $K$, $\eta$). Using the combined constraints from masses and decay patterns we propose spectroscopic assignments for several recently observed resonances, including $D_2(2740)^{0}$, $D^*_3(2750)$, $D_0(2550)^{0}$, $D^*_1(2600)^{0}$, $D_1^*(2760)^0$, $D^*_J(3000)$, $D_J(3000)$, $D^*_2(3000)$, $D^*_{s1}(2860)^{\pm}$ and $D^*_{s3}(2860)^{\pm}$. We also identify $D_{sJ}(3040)^+$ as a $2P$ excitation with $J^P=1^+$.}
Extending our model, we also calculate the mass spectra of doubly charmed $\Xi_{cc}$ and $\Omega_{cc}$ baryons within the heavy-diquark–light-quark picture. These theoretical predictions provide crucial guidance for ongoing and future experimental searches for higher radial and orbital excitations in the charmed meson and doubly charmed baryon sectors.

\end{abstract}
	
\maketitle
	
\section{INTRODUCTION}

\begin{table}[h]
	\centering
		\caption{\label{1}Masses and spin-parity ($J^{P}$) quantum numbers of the experimentally observed $D$ and $D_s$ meson and $\Xi_{cc}$ baryon states, as reported by the Particle Data Group (PDG) \cite{PDG2024}. The states marked as $UP$ and $NP$ indicate mesons with unnatural parity and natural parity, respectively.}
	\begin{ruledtabular}
	\begin{tabular}{lll}
		State & {$J^{P}$} & Mass (MeV) \\	
		\hline
		$D^{\pm}$               & $0^{-}$ & 1869.66$\pm$0.05\\
		$D^{0}$                 & $0^{-}$ & 1864.84$\pm$0.05\\
		$D^{*}(2007)^{0}$       & $1^{-}$ & 2006.85$\pm$0.05\\
		$D^{*}(2010)^{\pm}$     & $1^{-}$ & 2010.26$\pm$0.05\\
		$D_{0}^{*}(2300)$       & $0^{+}$ & 2343$\pm$10\\
		$D_{1}(2420)$           & $1^{+}$ & 2422.1$\pm$0.6\\
		$D_{1}(2430)^{0}$       & $1^{+}$ & 2412$\pm$9\\
		$D_{2}^{*}(2460)$       & $2^{+}$ & 2461.1$\pm$0.8\\
		$D_{0}(2550)^{0}$       & $0^{-}$ & 2549$\pm$19\\
		$D_{1}^{*}(2600)^{0}$   & $1^{-}$ & 2627$\pm$10\\
		$D_{2}(2740)^{0}$       & $2^{-}$ & 2747$\pm$6\\
		$D_{3}^{*}(2750)$       & $3^{-}$ & 2763.1$\pm$3.2\\
		$D_{1}^{*}(2760)^{0}$   & $1^{-}$ & 2781$\pm$22\\
		$D_{J}^{*}(3000)^{0}$           & $NP$ &3008.1$\pm$4.0 \\
		$D_{J}(3000)^{0}$           & $UP$ & 2971.8$\pm$8.7\\
		$D_{2}^{*}(3000)$           & $2^{+}$ & 3214$\pm$29$\pm$49\\
		\hline
		$D_s^{\pm}$                    & $0^{-}$ & 1968.36$\pm$0.07\\
		$D_s^{*\pm}$                   & $1^{-}$ & 2112.2$\pm$0.4\\
		$D_{s0}^{*}(2317)^{\pm}$       & $0^{+}$ & 2317.8$\pm$0.5\\
		$D_{s1}(2460)^{\pm}$           & $1^{+}$ &2459.5 $\pm$0.6\\
		$D_{s1}(2536)^{\pm}$           & $1^{+}$ & 2535.11$\pm$0.06\\
		$D_{s2}^{*}(2573)$             & $2^{+}$ & 2569.1$\pm$0.8\\
		$D_{s0}(2590)^{+}$             & $0^{-}$ & 2591$\pm$9\\
		$D_{s1}^{*}(2700)^{\pm}$       & $1^{-}$ & 2714$\pm$5\\
		$D_{s1}^{*}(2860)^{\pm}$       & $1^{-}$ & 2859$\pm$27\\
		$D_{s3}^{*}(2860)^{\pm}$       & $3^{-}$ & 2860$\pm$7\\
		$D_{sJ}(3040)^{\pm}$           & $?^{?}$ & 3004$^{+31}_{-9}$\\
		\hline
		$\Xi_{cc}^{+}$           & $?^{?}$ & 3518.9$\pm$0.9\\
		$\Xi_{cc}^{++}$           & $?^{?}$ & 3621.6$\pm$0.4\\
	\end{tabular}
	\end{ruledtabular}
\end{table}

The spectroscopy of open-charm hadrons, including charmed mesons ($D$ and $D_s$) and doubly charmed baryons ($\Xi_{cc}$ and $\Omega_{cc}$), provides a crucial window into the nonperturbative dynamics of Quantum Chromodynamics (QCD). These systems involve a heavy charm quark bound to light quark constituents, and thus sit at the nexus of heavy-quark physics and light-quark dynamics. 

Over the past two decades, significant experimental progress has been made in identifying numerous charmed meson states, primarily through observations by $B$-factories (\textit{BABAR}, Belle), CLEO, and LHCb collaborations. 
Consequently, many $D$ and $D_s$ resonances are now listed by the Particle Gata Group (PDG) \cite{PDG2024} as shown in Table [\ref{1}]. As new states are discovered, mapping out the mass spectra of $D$ and $D_s$ mesons becomes essential to distinguish conventional mesonic states from possible exotic configurations and to classify conventional mesonic states within the quark model framework. Furthermore, the spin-parity quantum numbers assignment for $D_{J}^{*}(3000)^{0}$, $D_{J}(3000)^{0}$, $D_{sJ}(3040)^{\pm}$ states have not been firmly established by experiment.

On the baryon side, the doubly charmed baryons $\Xi_{cc}$ and $\Omega_{cc}$ (with quark content $ccq$ and $ccs$, respectively) represent an important frontier in heavy hadron physics. 
Experimentally, evidence for the $\Xi_{cc}$ was first claimed by the SELEX experiment in 2002, which reported a $\Xi_{cc}^+$ state at a mass of about 3519~MeV. However, subsequent searches by BaBar, Belle, and LHCb failed to confirm this signal. The breakthrough came in 2017 when the LHCb Collaboration unambiguously observed the $\Xi_{cc}^{++}$ baryon in the $\Lambda_c^+ K^- \pi^+ \pi^+$ channel, measuring its mass to be about 3621~MeV \cite{PDG2024}. This state is most naturally interpreted as the ground-state $\Xi_{cc}$ with quantum numbers $J^P=\tfrac{1}{2}^+$. LHCb has since measured the $\Xi_{cc}^{++}$ mass with high precision \cite{ PDG2024}. To date, no $\Omega_{cc}$ (the $ccs$ baryon) has been observed in experiment. The spin-parity of the $\Xi_{cc}^{++}$ is yet to be directly measured; establishing $J^P$ and discovering excited states of $\Xi_{cc}$ and $\Omega_{cc}$ are among the next objectives for experiments like LHCb. 

Over the past two decades, various theoretical models have been developed to study the spectroscopy of charmed mesons and baryons. Constituent quark models—both nonrelativistic \cite{Li:2010vx, Segovia:2015dia, Kher:2017wsq, Ni:2021pce, Valcarce:2008dr, Giannuzzi:2009gh, Roberts:2007ni, Albertus:2006ya, Shah:2016vmd, Shah:2017liu} and relativistic  \cite{ Godfrey:1985xj, Ebert:2009ua, Devlani:2011zz, Liu:2013maa, Badalian:2011tb,Sun:2013qca, Devlani:2013kta, Lu:2014zua, Song:2015fha, Song:2015nia, Godfrey:2015dva, Kher:2017wsq, Ebert:2002ig,Yoshida:2015tia, Karliner:2014gca}—remain central to this effort. These treat mesons as \( q\bar{q} \) and baryons as \( qqq \) or quark–diquark systems with inter-quark potentials inspired by QCD, typically incorporating one-gluon exchange and a confining term. Relativized models such as the Godfrey–Isgur model \cite{Godfrey:1985xj} and its extensions have reliably predicted the spectra of charmed mesons, while modern relativistic quark models continue to refine the spectrum of excited \( D \) and \( D_s \) states \cite{ Ebert:2009ua,Godfrey:2015dva}. In the baryon sector, the relativistic quark model by Ebert el al. \cite{Ebert:2002ig} has accurately computed ground state mass of \( \Xi_{cc} \), before its experimental confirmation.

Further, various theoretical methods have been put forward to study charmed mesons and baryons, including  Bohr–Sommerfeld quantization approach \cite{Chen:2018nnr}, QCD sum rule \cite{Zhou:2014ytp, Gelhausen:2014jea, Zhang:2008rt, Wang:2010hs}, $^{3}P_{0}$ pair-creation model \cite{Ferretti:2015rsa}, heavy quark effective theory \cite{Korner:1994nh}, the extended Linear Sigma Model \cite{Eshraim:2014eka}, a relativistic Dirac formalism \cite{Shah:2014caa}, lattice QCD \cite{Kawanai:2015tga, Nochi:2016wqg, Mohler:2011ke, Cheung:2016bym, Liu:2009jc, Alexandrou:2012xk, Padmanath:2015jea}, Regge phenomenology \cite{Oudichhya:2022ssc}, effective field theory \cite{Soto:2020pfa}. 

Although various theoretical and experimental investigations have been conducted, the spin-parity quantum number assignments for states such as $D_{J}^{*}(3000)^{0}$, $D_{J}(3000)^{0}$, and $D_{sJ}(3040)^{\pm}$ remain unresolved, and even the ground states of doubly heavy baryons have yet to be established. The primary objective of the present study is to deliver a comprehensive theoretical account of the mass spectra of open-charm hadrons—specifically, the $D$ and $D_s$ mesons and the $\Xi_{cc}$ and $\Omega_{cc}$ baryons, upto higher orbital and radial excitations within the framework of the relativistic flux-tube model.

Earlier studies employing the flux tube model, such as Ref. \cite{Jia:2018vwl}, calculated only spin-averaged masses for the $D$ and $D_s$ mesons. However, their predictions could not discriminate between spin multiplets because spin-dependent interactions were not considered. Additionally, in Ref. \cite{Chen:2009zt, Shan:2008ga}, calculations were limited to the $1P$ and $1D$ states for the $D$ and $D_s$ mesons, neglecting higher radial and orbital excitations. Similarly, the authors in Ref. \cite{jia2023} restricted their calculations to a few low-lying states of the $\Xi_{cc}$ and $\Omega_{cc}$ baryons.

In the current study, we employ the relativistic flux-tube derived mass formula to compute spin-averaged masses. Crucially, our methodology integrates spin-dependent interactions, enabling predictions of hyperfine structure  splittings within a $jj$ coupling scheme suitable for heavy-light systems. We aim to identify the masses of fairly high orbital and radial excited states. This approach has been validated in our previous research, successfully reproducing mass spectra for singly heavy baryons \cite{Jakhad:2023ids}. In the present work, we extend the application of this model to open-charm mesons and doubly charmed baryons. Based on the predicted mass spectra, we further propose tentative spin-parity ($J^P$) assignments for several experimentally observed charmed meson states whose quantum numbers remain unconfirmed. Additionally, we provide theoretical predictions for the excitation spectrum of the doubly charmed baryons $\Xi_{cc}$ and $\Omega_{cc}$, which may serve as valuable input for future experimental searches aimed at discovering and classifying their excited states.

The structure of the paper is organized as follows. In Section~II, we outline the relativistic flux-tube model as applied to  charmed meson and doubly charmed baryon and detail the methodology used to compute their mass spectra. {In Section~III, we present the framework used to evaluate the strong decay widths.} Section~IV contains our numerical results for the spectra and decay widths, along with a systematic comparison to available experimental measurements and existing theoretical studies. Finally, Section~IV summarizes our main conclusions.

\section{MASS SPECTRUM}

\subsection{Spin-averaged mass in relativistic flux tube model}

We consider the charmed meson or baryon as a two-body system consisting of a heavy component and a light component are bounded by confining color field which behaves like a  straight, rotating  flux tube (string). This flux tube carries a constant tension $\sigma$ (analogous to a string’s energy per unit length) which account for the linear rising potential at large separations in QCD dynamics.

In natural units ($c=1$), the system’s dynamics can be derived from the Lagrangian of a rotating relativistic string with endpoint masses. In the rigid-rotation approximation (assuming a constant angular velocity $\omega$ about the center-of-mass), the Lagrangian takes the form \cite{LaCourse:1988cu, Jakhad:2023ids}

\begin{equation}
\label{eq1}
\mathscr{L} \;=\; \sum_{i} m_{0_i}\,\sqrt{\,1 - \omega^2 r_i^2\,}\;+\; T \sum_{i}\int_{0}^{\,r_i} dr\,\sqrt{\,1 - \omega^2 r^2\,}\,.
\end{equation}

Here $i=l,h$ label the light and heavy components, with mass $ m_{0_l}$ and $ m_{0_h}$, respectively. The quantity $r_i$ is the distance of mass $ m_{0_i}$ from the center of mass of system. 

From the Lagrangian, one can derive the total orbital angular momentum $L$ using $ L= \partial \mathscr{L}/\partial \omega$  as

\begin{equation}
\label{eq2}
L =\sum_{i} \left[ \frac{m_{0_i}\,v_i^2}{\,\omega\sqrt{\,1-v_i^2\,}\,} + \frac{T}{2\,\omega^2}\Big(\sin^{-1} v_i - v_i\sqrt{\,1-v_i^2\,}\Big)\right].
\end{equation}

The first term inside the sum is the relativistic angular momentum of each orbiting mass. The second term is the contribution of the rotating string, obtained by integrating the differential string angular momentum density.

The total relativistic energy (Hamiltonian) of the system $H = (\omega \partial \mathscr{L}/\partial \omega) - \mathscr{L}$ , which corresponds to the spin-averaged hadron mass $\bar{M}$, is derived from the Lagrangian  as

\begin{equation}
\label{eq3}
\bar{M} \;=\; \sum_{i} \frac{m_{0_i}}{\sqrt{\,1-v_i^2\,}} \;+\; \frac{T}{\omega}\Big(\sin^{-1} v_l + \sin^{-1} v_h\Big) \,. 
\end{equation}

It is often convenient to define dynamical masses for the rotating   light and heavy objects as $ m_l = m_{0_l}/\sqrt{1-v_l^2}$ and $ m_h = m_{0_h}/\sqrt{1-v_h^2}$, respectively. Then we have

\begin{align}
	\label{eq4}
	L = \frac{1}{\omega}(m_{l} v_{l}^{2} + m_{h} v_{h}^{2})
	&+ \frac{T}{2 \omega^{2}} \Big( 
	\sin^{-1} v_{l} - v_{l} \sqrt{1 - v_{l}^{2}} \notag \\
	&\quad + \sin^{-1} v_{h} - v_{h} \sqrt{1 - v_{h}^{2}} 
	\Big),
\end{align}
and 
\begin{equation}
\label{eq5}
\bar{M} =  m_h +  m_l + \frac{T}{\omega}(\sin^{-1} v_h + \sin^{-1} v_l).
\end{equation}

Eq. [\ref{eq3}] and Eq. [\ref{eq4}] together define $L$ and $\bar{M}$ as functions of the unknown angular velocity $\omega$. To find a direct relation between $\bar{M}$ and $L$, we eliminate $\omega$ by using the equilibrium conditions and making a physically motivated approximation. 

The equilibrium of the rotating system requires that the tension $T$ in the string provides the necessary centripetal force on each mass. In particular, for the heavy end we have

\begin{equation}
	\label{eq6} 
T\;=\frac{m_h v_h \omega}{\sqrt{1 - v_{h}^{2}}}=m_h v_h \omega [1+\frac{v_h^2}{2}+\frac{3v_h^4}{8}+...]\simeq m_h v_h \omega,  
\end{equation}
to leading order in $v_h$ which is consistent with the small-$v_h$ limit.
 
For heavy–light mesons and baryons, the flux tube model must address a highly asymmetric scenario. In this case, one end of the flux tube connects to a heavy quark, nearly static due to its large mass, while the other end attaches to a significantly lighter quark that demands a fully relativistic treatment ($m_h \gg m_l$). In this regime the light component is highly relativistic, with $v_l$ approaching 1, while the heavy component moves more slowly ($v_h \ll 1$). 
 
In this heavy–light limit, we expand the spin-averaged mass and angular momentum expressions, Eq.[\ref{eq4}] and [\ref{eq5}], up to second order in the small parameter $v_h$. Subsequently, using the relation $T/\omega \approx m_h v_h$ derived from Eq.[\ref{eq6}] to eliminate $\omega$, we ultimately obtain a clear, Regge-like relationship between the spin-averaged mass and angular momentum \cite{article,Jakhad:2023ids}:

\begin{equation}
\label{eq7}
(\bar{M} - m_h)^2 = \sigma\,\frac{L}{2} \;+\; (m_l + m_h v_h^2)^2 
\end{equation}

The relationship between the angular speed $\omega$ of the rotating flux tube and the orbital angular momentum $L$ can be obtained by combining Eqs.~[\ref{eq4}] and [\ref{eq7}] as \cite{article,Jakhad:2023ids}
\begin{equation}
	\label{eq8}
	\omega = \sqrt{\frac{\sigma}{8L}}.
\end{equation}
Using this result, we can then express the distance between the heavy and light constituents as \cite{article, Jakhad:2023ids}
\begin{equation}
	\label{eq9}
	r = \frac{v_l + v_h}{\omega} = (v_l + v_h)\sqrt{\frac{8L}{\sigma}}.
\end{equation}
	
So far we have considered only the leading (lowest radial) trajectory in $(L,\,M^2)$ plane. In reality, hadrons can also be excited radially. A striking empirical fact, supported by quantum mechanical approach \cite{Olson:1993ux,Allen:1999dk}, is that radial excitations of a heavy-light meson tend to lie on parallel Regge trajectories in the $(L,\,M^2)$ plane (i.e. daughters roughly parallel to the leading trajectory). In the context of our semiclassical model, this can be accommodated by a simple modification of Eq. [\ref{eq7}]. Following Refs. \cite{Olson:1993ux, Allen:1999dk}, we introduce a phenomenological offset to $L$ for each radial quantum number. In practice, one replaces $L$ by $L + \lambda\,n_r$ in the Regge formula. Here, $n_r=n-1$ (with $n=1,2,...)$ is the radial excitation quantum number (with $n_r=0$ for the ground state, $n_r=1$ for the first radial excitation, etc.), and $\lambda$ is a constant offset parameter fitted to the spacing between radial trajectories. The modified  relations are \cite{article,Jakhad:2023ids}
\begin{equation}
	\label{eq10}
\big(\bar{M} - m_h\big)^2 \;=\; \frac{\sigma}{2}\,\Big(L + \lambda\,n_r\Big) \;+\; \big(m_l + m_h v_h^2\big)^2 \, ,
\end{equation} 
\begin{equation}
	\label{eq11}
	r = (v_l + v_h)\sqrt{\frac{8(L + \lambda\,n_r)}{\sigma}}.
\end{equation}

By construction, all states with different $n_r$ (but same $m_h, m_l$) now lie on parallel lines of slope $\sigma/2$ in the $(L,(\bar{M}-m_h)^2)$ plane, separated by the constant spacing $\Delta (\bar{M}-m_h)^2 = \frac{\sigma}{2}\lambda$ between successive $n_r$ levels. This extended formula allows one to estimate the masses of both orbital ($L$) and radial ($n_r$) excitations in the heavy-light system.
The parameter $\lambda$ is treated as universal for a given class of hadrons (it can be extracted by fitting the known spectrum).

\subsection{Spin-dependent corrections}

The flux-tube model described so far treats the quarks (or diquarks) as spinless, so $\bar{M}$ corresponds to a spin-averaged mass. To obtain the physical  masses, we must include spin-dependent interactions that split the degeneracies among states of different total spin $J$ (for a given $L,n_r$). We incorporate these effects through a perturbative Hamiltonian as 

\begin{equation}
\Delta M \;=\; H_{so} \;+\; H_{t} \;+\; H_{ss}\,.
\end{equation}

The spin–orbit Interaction term ($H_{so}$) arises from the combination of short-range color magnetic forces (one-gluon exchange) and long-range Thomas precession. It can be written as an operator proportional to the orbital angular momentum ($\mathbf{L}$) dotted into the spin of the light component ($\mathbf{S}_l$) and of the heavy component ($\mathbf{S}_h$) in the form 
\begin{equation} \label{eq10.1}
H_{so} = a_1\,\mathbf{L}\cdot\mathbf{S}_l + a_2\,\mathbf{L}\cdot\mathbf{S}_h.
\end{equation}
The coefficients $a_1$ and $a_2$ are
\begin{equation}
a_1 \;=\; \Big(\frac{2\alpha}{3r^3} - \frac{b_0}{2r}\Big)\frac{1}{m_l^2}\;+\;\frac{4\alpha}{3r^3}\frac{1}{m_h m_l},
\end{equation}
\begin{equation}
a_2 \;=\; \Big(\frac{2\alpha}{3r^3} - \frac{b_0}{2r}\Big)\frac{1}{m_h^2}\;+\;\frac{4\alpha}{3r^3}\frac{1}{m_h m_l}. 	 	
\end{equation}
Here $\alpha$ is the coupling constant of the one-gluon-exchange potential, and $b_0$ is a parameter associated with the long-range confining potential. 

The tensor interaction term ($H_t$)  comes from the long-range part of the color-magnetic dipole–dipole interaction between the spins. Its operator form is
\begin{equation}\label{eq10.2}
	H_t = \frac{4\alpha}{3 r^{3} m_h m_l}\mathbf{\hat{B}}.
\end{equation}
Here, $\mathbf{\hat{B}}$ is standard tensor operator with form
\begin{equation}
 \mathbf{\hat{B}} =\frac{ 3(\mathbf{S}_l \cdot \mathbf{r})(\mathbf{S}_h \cdot \mathbf{r}) }{r^{2}}- \mathbf{S}_l \cdot \mathbf{S}_h.
 \end{equation}

The spin–spin contact interaction term ($H_{ss}$) is a short-range hyperfine term. We take a contact form
\begin{equation} \label{eq10.3}
H_{ss}=c\,\mathbf{S}_l\cdot \mathbf{S}_h;	
\end{equation}
often modeled by a smeared delta-function interaction in coordinate space. For instance, a Gaussian-smeared contact with strength $c= (32\pi \alpha \sigma_0^3)/(9\sqrt{\pi}\,m_h m_l)$ and width parameter $\sigma_0$ can be used. In heavy–light mesons, $H_{ss}$ gives the well-known hyperfine splitting between (spin-weighted) spin-singlet and spin-triplet configurations. A familiar example is the mass difference between a pseudoscalar meson (total $S_{tot}=0$) and its vector meson partner ($S_{tot}=1$). In our model, the constants $b_0$ and $\sigma_0$ (as well as $\alpha$) are chosen to reproduce known splittings in the charmed meson spectrum, and are kept fixed for predictions. 

With these spin-dependent terms, the total mass for the heavy light system is specified as $M=\bar{M}+\Delta M$. We adopt the $j$–$j$ coupling scheme appropriate for heavy–light hadrons to calculate the expectation value of $\Delta M$ in $|J,j\rangle$ basis. For singly charmed mesons, one typically couples the light antiquark spin $\mathbf{S}_{l}$ with the orbital $\mathbf{L}$ to form an intermediate $\mathbf{j} = \mathbf{L} + \mathbf{S}_l$ (sometimes called the total angular momentum of the light cloud). This $\mathbf{j}$ is then coupled with the heavy quark spin $\mathbf{S}_h $ to give the total $\mathbf{J}$.
For doubly charmed baryons in the quark–diquark picture, we do the analogous coupling: the light quark spin $\mathbf{S}_l $ is coupled with $\mathbf{L}$ to produce an intermediate $\mathbf{j}$. 
Subsequently, $\mathbf{j}$ couples with the diquark spin $\mathbf{S}_h $ to yield the baryon’s $\mathbf{J}$. 

This scheme is particularly convenient because in the limit $m_h\to \infty$ the heavy quark's spin decouples, making $j$ an approximate conserved quantum number (as in heavy quark effective theory). 
We calculate the expectation values of the operators $\mathbf{L}\cdot \mathbf{S}_h$, $\mathbf{L}\cdot \mathbf{S}_l$, $\hat{\mathbf{B}}$, and $\mathbf{S}_h\cdot \mathbf{S}_l$ for specific $(L,J,j)$ configurations corresponding to singly charmed mesons and doubly charmed baryons. These computed values are presented in Tables (\ref{table1}) and (\ref{table2}), respectively. A detailed derivation of these results is provided in the Appendix for reader new to research in this field.

	\begin{table}
	\caption{\label{table1}Expectation values of spin-dependent operators for various $(L, J, j)$ configurations in heavy–light mesons.}
	\begin{ruledtabular}
		\begin{tabular}{ccccccc}
			\multicolumn{3}{c}{Quantum Numbers} &  &  &  &  \\
			\cline{1-3}
			$L$ & $j$ &$J$ &  $\;\langle\mathbf{L}\!\cdot\!\mathbf{S}_l\rangle\;$ & $\;\langle\mathbf{L}\!\cdot\!\mathbf{S}_h\rangle\;$ & $\;\langle\mathbf{\hat{B}}\rangle\;$ & $\;\langle\mathbf{S}_h\!\cdot\!\mathbf{S}_l\rangle\;$ \\
			\hline
			0 & 1/2  & 0   &   0   &    0    &   0   &  –3/4  \\
			0 & 1/2  & 1   &   0   &    0    &   0   &   1/4  \\
			1 & 1/2  & 0   &  –1   &   –1    &  –1   &   1/4  \\
			1 & 1/2  & 1   &  –1   &  1/3    &  1/3  & –1/12  \\
			1 & 3/2  & 1   &  1/2  &  –5/6   &  1/6  & –5/12  \\
			1 & 3/2  & 2   &  1/2  &   1/2   & –1/10 &   1/4  \\
			2 & 3/2  & 1   & –3/2  &  –3/2   & –1/2  &   1/4  \\
			2 & 3/2  & 2   & –3/2  &   9/10  &  3/10 & –3/20  \\
			2 & 5/2  & 2   &   1   &  –7/5   &  1/5  & –7/20  \\
			2 & 5/2  & 3   &   1   &    1    & –1/7  &   1/4  \\
			3 & 5/2  & 2   &  –2   &   –2    & –2/5  &   1/4  \\
			3 & 5/2  & 3   &  –2   &  10/7   &  2/7  & –5/28  \\
			3 & 7/2  & 3   &  3/2  & –27/14  &  3/14 & –9/28  \\
			3 & 5/2  & 4   &  3/2  &   3/2   & –1/6  &   1/4  \\
			4 & 7/2  & 3   & –5/2  &  –5/2   & –5/14 &   1/4  \\
			4 & 7/2  & 4   & –5/2  &  35/18  &  5/18 & –7/36  \\
			4 & 9/2  & 4   &   2   &  –22/9  &  2/9  & –11/36 \\
			4 & 9/2  & 5   &   2   &    2    & –2/11 &   1/4  \\
			5 & 9/2  & 4   &  –3   &   –3    & –1/3  &   1/4  \\
			5 & 9/2  & 5   &  –3   &  27/11  &  3/11 & –9/44  \\
			5 & 11/2 & 5   &  5/2  & –65/22  &  5/22 & –13/44 \\
			5 & 11/2 & 6   &  5/2  &   5/2   & –5/26 &   1/4  \\
		\end{tabular}
	\end{ruledtabular}
\end{table}

\begin{table}
	\caption{\label{table2} Expectation values of spin-dependent operators for various $(L, J, j)$ configurations in doubly heavy baryon with axial vector diquark ($S_{h}=1$).}
	\begin{ruledtabular}
		\begin{tabular}{ccccccc}
			\multicolumn{3}{c}{Quantum Numbers} &  &  &  &  \\
			\cline{1-3}
			$L$ & $j$ & $J$ &  $\;\langle\mathbf{L}\!\cdot\!\mathbf{S}_l\rangle\;$ & $\;\langle\mathbf{L}\!\cdot\!\mathbf{S}_h\rangle\;$ & $\;\langle\mathbf{\hat{B}}\rangle\;$ & $\;\langle\mathbf{S}_h\!\cdot\!\mathbf{S}_l\rangle\;$ \\
			\hline
			0 & 1/2 & 1/2  &   0   &    0   &    0   &  –1    \\
			0 & 1/2 & 3/2  &   0   &    0   &    0   &  1/2   \\
			1 & 1/2 & 1/2  &  –1   & –4/3   & –4/3   &  1/3   \\
			1 & 3/2 & 1/2  &  1/2  & –5/3   &  1/3   & –5/6   \\
			1 & 1/2 & 3/2  &  –1   &  2/3   &  2/3   & –1/6   \\
			1 & 3/2 & 3/2  &  1/2  & –2/3   &  2/15  & –1/3   \\
			1 & 3/2 & 5/2  &  1/2  &   1    & –1/5   &  1/2   \\
			2 & 3/2 & 1/2  & –3/2  &   –3   & –1     &  1/2   \\
			2 & 3/2 & 3/2  & –3/2  &  –6/5  & –2/5   &  1/5   \\
			2 & 5/2 & 3/2  &   1   & –14/5  &  2/5   & –7/10  \\
			2 & 3/2 & 5/2  & –3/2  &   9/5  &  3/5   & –3/10  \\
			2 & 5/2 & 5/2  &   1   &  –4/5  &  4/35  & –1/5   \\
			2 & 5/2 & 7/2  &   1   &    2   & –2/7   &  1/2   \\
			3 & 5/2 & 1/2  &  –2   &   –4   & –4/5   &  1/2   \\
			3 & 5/2 & 5/2  &  –2   &  –8/7  & –8/35  &  1/7   \\
			3 & 7/2 & 5/2  &  3/2  & –27/7  &  3/7   & –9/14  \\
			3 & 5/2 & 7/2  &  –2   &  20/7  &  4/7   & –5/14  \\
			3 & 7/2 & 7/2  &  3/2  &  –6/7  &  2/21  & –1/7   \\
			3 & 7/2 & 9/2  &  3/2  &    3   & –1/3   &  1/2   \\
			4 & 7/2 & 5/2  & –5/2  &   –5   & –5/7   &  1/2   \\
			4 & 7/2 & 7/2  & –5/2  & –70/63 & –10/63 &  7/63  \\
			4 & 9/2 & 7/2  &   2   & –44/9  &  4/9   & –11/18 \\
			4 & 7/2 & 9/2  & –5/2  &  35/9  &  5/9   & –7/18  \\
			4 & 9/2 & 9/2  &   2   &  –8/9  &  8/99  & –1/9   \\
			4 & 9/2 & 11/2 &   2   &    4   & –4/11  &  1/2   \\
		\end{tabular}
	\end{ruledtabular}
\end{table}

\subsection{Estimation of model parameters and associated uncertainties} 

\subsubsection{$D$ and $D_s$ mesons}
To estimate the optimal values of the model parameters and their associated uncertainties, we utilized the iminuit package in Python \cite{iminuit2021}, which is based on the MINUIT algorithm developed at CERN for robust function minimization \cite{minuit1975}. 
The fitting was performed using the least-squares method by minimizing the $\chi^2$ difference between the theoretical  mass predictions by our model and the experimentally measured masses of $D$ and $D_s$ mesons, including both ground and excited states. 
The experimental measured masses, along with their respective uncertainties, of  the well-established states $D^{\pm}$, $D^{0}$, $D^*(2007)^0$, $D^*(2010)^\pm$, $D_0^{*}(2300)$, $D_1(2420)$, $D_1(2430)^0$, $D_2^{*}(2460)$, $D_3^{*}(2750)$, $D_s^{\pm}$, $D_s^{*\pm}$, $D_{s1}(2536)^{\pm}$, $D_{s2}^{*}(2573)$, $D_{s1}^{*}(2700)^{\pm}$, $D_{s3}^{*}(2860)^{\pm}$, and mass of charm quark $m_{0_{c}}=1273\pm4.6~\text{GeV}$,  are used as inputs \cite{PDG2024}.
The parameters values along with their statistical uncertainties that correspond to the best agreement between the theoretical masses and the experimentally measured masses are:
the mass of the $c$ quark $m_c = 1375.68 \pm 0.15~\text{MeV}$, 
the mass of $u$ or $d$ quark $m_{u/d} = 399.95 \pm 0.02~\text{MeV}$, 
the mass of the $s$ quark $m_s = 484.29 \pm 0.06~\text{MeV}$, 
the string tension for $D$ meson $\sigma_D = (1.5321 \pm 0.0023) \times 10^6~\text{MeV}^2$, 
the string tension for $D_{s}$ meson $\sigma_{D_{s}} = (1.7882 \pm 0.0016) \times 10^{6}~\text{MeV}^2$, 
$\lambda = 1.479 \pm 0.015$, 
$\alpha = 0.306 \pm 0.005$, $b_{0} = (36.2 \pm 1.2) \times 10^3~\text{MeV}^2$, and 
$\sigma_0 = 502.8 \pm 2.8~\text{MeV}$.


\subsubsection{$\Xi_{cc}$ and $\Omega_{cc}$ baryons with the diquark in the ground state}

In $\Xi_{cc}$ and $\Omega_{cc}$ baryons, the two $c$ quarks form a diquark in the internal ($1S$) ground state. By Fermi statistics (Pauli exclusion), the spatially symmetric configuration forces a symmetric spin wave function, so the diquark is an axial-vector state, denoted \(\{cc\}\), with heavy-diquark spin ($S_h=1$).

The mass of a heavy diquark ($m_{0_{\{cc\}}}$) is approximated by the sum of the individual charm quark mass: $m_{0_{\{cc\}}} = 2m_{0_{c}} = 2546\pm5.6~\text{MeV}$ \cite{PDG2024}.

Due to the limited experimental data on the masses of doubly charmed baryons—aside from the well-established state $\Xi_{cc}^{++}$—it is necessary to rely on theoretical mass predictions for $1S-$wave states to extract the model parameters. The study by Ebert et al. \cite{Ebert:2002ig} successfully predicted the ground state mass of the $\Xi_{cc}$ even prior to its experimental discovery, estimating it at $3620~\text{MeV}$. This prediction closely matches the experimentally measured mass of $3621.55,\text{MeV}$ reported by the LHCb collaboration \cite{PDG2024}. Hence take predictions for the $1S-$ wave states for $\Xi_{cc}$ and $\Omega_{cc}$ baryons as input from ref. \cite{Ebert:2002ig} to determine the dynamical mass of the diquark $m_{\{cc\} }= 2802.46~\text{MeV}$ and $\alpha = 0.309$.

In the heavy quark limit ($m_{h}$$\rightarrow$$\infty$), two heavy quarks inside a doubly heavy diquark behave analogously to a heavy antiquark with respect to a light quark \cite{PhysRevD.73.054003}. Consequently, the light degrees of freedom in a doubly heavy baryon experience the same color field as in a heavy-light meson. Due to the identical color field structures, heavy quark-diquark symmetry at leading order implies identical string tensions ($\sigma$) for these systems. However, finite mass effects, arising when heavy quark masses deviate from infinity, introduce small symmetry violations. Such finite mass corrections can be effectively parameterized through empirical power-law scaling relations of the string tensions in terms of the heavy quark masses \cite{jia2023}:

\begin{equation}
	\frac{\sigma_{D}}{\sigma_{B}} = \left(\frac{m_c}{m_b}\right)^P, \quad
	\frac{\sigma_{D}}{\sigma_{\Xi_{cc}}} = \left(\frac{m_c}{m_{\{cc\}}}\right)^P.
	\label{eq:string_scaling_non_strange}
\end{equation}

Extending this phenomenological approach to strange heavy-light hadrons, we have a similar relation \cite{jia2023}:
\begin{equation}
	\frac{\sigma_{D_{s}}}{\sigma_{B_{s}}} = \left(\frac{m_c}{m_b}\right)^Q, \quad
	\frac{\sigma_{D_{s}}}{\sigma_{\Omega_{cc}}} = \left(\frac{m_c}{m_{\{cc\}}}\right)^Q.
	\label{eq:string_scaling_strange}
\end{equation}

Using the known string tensions for the \( D \)-, \( B \)-, \( D_s \)-, and \( B_s \)-mesons, along with the quark masses \( m_c \), \( m_b \), and the diquark mass \( m_{\{cc\}} \) fitted from well-established experimental states, we extract the scaling parameters \( P=0.1157\pm0.002 \) and \( Q=0.0927\pm0.0017\), and subsequently determine the string tensions \( \sigma_{\Xi_{cc}} = (1.6636 \pm 0.0034) \times 10^6~\text{MeV}^2 \) and \( \sigma_{\Omega_{cc}} = (1.9101 \pm 0.0029) \times 10^6~\text{MeV}^2 \).

\subsubsection{{$\Xi_{cc}$ and $\Omega_{cc}$ baryons with the diquark in an excited state}}

{In this subsection we analyze those doubly heavy configurations in which the two charm quarks inside the diquark are themselves excited. For internal excitations of the \(cc\) subsystem, the heavy diquark is required to appear in a spin–singlet state (\(S_h = 0\)) for the \(1P\) and \(2P\) waves, whereas the internal \(2S\) excitation corresponds to a spin–triplet diquark (\(S_h = 1\)).}

{The masses of the \(\Xi_{cc}\) and \(\Omega_{cc}\) baryons, in which the light quark is orbitally or radially excited with respect to such an internally excited heavy diquark in the \(2S\), \(1P\), and \(2P\) waves, can be obtained from  Eqs. [\ref{eq10}, \ref{eq10.1}, \ref{eq10.2}, \ref{eq10.3}]. To make use of these relations, we first determine the masses of the heavy diquarks,
\(
m_{h_{2S}}, m_{h_{1P}}, m_{h_{2P}},
\)
corresponding to the internal \(2S\), \(1P\), and \(2P\) excitations, respectively.
}
{As a starting point, we consider the charmonium system within the relativistic flux tube framework. A charm quark and an anticharm quark of mass \(m_c\), connected by a stringlike flux tube and carrying relative orbital angular momentum \(L\) and principal quantum number \(n = n_r + 1\), satisfy a Regge-type relation for the spin-averaged mass \(\bar M\):}

\begin{equation}
	\label{eq3.1}
	\bar{M} = 2m_c +  \left( \frac{\sigma_{c\bar{c}}^2}{2 \pi^{2} m_c} \right)^{\tfrac{1}{3}}(\lambda n_r + L)^{\frac{2}{3}},
\end{equation}
{where $\sigma_{c\bar{c}}$ denotes the charmonium string tension. From this expression, a direct relationship between the spin-averaged masses of the $1S$ and $1P$ charmonium waves can be derived:
}\begin{equation}
	\bar{M}_{1P}= \bar{M}_{1S} + \left( \frac{\sigma_{c\bar{c}}^2}{\pi^{2}\bar{M}_{1S}} \right)^{\tfrac{1}{3}}
\end{equation}
{We utilize the experimentally measured masses of the $1S$ and $1P$ charmonium states---specifically $\eta_{c}(1S)$, $J/\psi(1S)$, $\chi_{c0}(1P)$, $\chi_{c1}(1P)$, $h_{c}(1P)$, and $\chi_{c2}(1P)$---to compute the corresponding spin-averaged masses for the $1S$ and $1P$ waves. This procedure allows us to extract the string tension for charmonium, yielding $\sigma_{c\bar{c}}=(1.698 \pm 0.0006) \times 10^6$ MeV$^2 $.}

{Building upon the established charmonium framework, we extend Eq.~(\ref{eq3.1}) to the color–antitriplet \(cc\) diquark. Denoting the principal quantum number of the diquark by \(n_d = n_{rd} + 1\) and its orbital angular momentum by \(L_d\), the analogous relation reads}
\begin{equation}
	\label{eq3.2}
	\bar{M} = 2m_c +\left( \frac{\sigma_{cc}^2}{2 \pi^{2}m_c} \right)^{\tfrac{1}{3}}(\lambda n_{rd} + L_d)^{\frac{2}{3}},
\end{equation}
{This leads to the following relations for the spin–averaged  mass of the diquark in $1P$, $2S$ and $2P$ waves: }
\begin{equation}
	\label{x1}
	\bar{M}_{1P} = \bar{M}_{1S} +  \left( \frac{\sigma_{cc}^2}{\pi^{2}\bar{M}_{1S}} \right)^{\tfrac{1}{3}}
\end{equation}
\begin{equation}
	\label{x2}
	\bar{M}_{2S} = \bar{M}_{1S} + \left( \frac{\sigma_{cc}^2}{\pi^{2}\bar{M}_{1S}} \right)^{\tfrac{1}{3}} \lambda^{\frac{2}{3}}
\end{equation}
\begin{equation}
	\label{x3}
	\bar{M}_{2P} = \bar{M}_{1S} +  \left( \frac{\sigma_{cc}^2}{\pi^{2}\bar{M}_{1S}} \right)^{\tfrac{1}{3}}(\lambda+1)^{\frac{2}{3}}
\end{equation}

{The physics governing the string tension $\sigma$ (related to the inverse Regge slope $1/\alpha'$ by $1/\alpha'\!\sim\! \sigma$) dictates its proportionality to the square root of the color factor, $1/\alpha' \propto \sqrt{C\,\alpha_s}$. Here, $C$ represents the quadratic Casimir of the color sources at the flux tube ends \cite{PhysRevD.13.1934}. For the $cc[\bar{\mathbf 3}_c]$ diquark (an antitriplet), $C=2/3$, whereas for the $c\bar c[\mathbf 1_c]$ meson (a color singlet), $C=4/3$. This implies a direct relationship between their respective string tensions:}
\begin{equation}
	\sigma_{\,cc}=\sqrt{\frac{2/3}{4/3}}\sigma_{\,c\bar c}
	=\frac{1}{\sqrt{2}}\sigma_{\,c\bar c}
\end{equation}
{Using this relationship and our previously determined value for $\sigma_{c\bar{c}}$, we find the string tension for the $cc$ diquark to be $\sigma_{\,cc}=(1.200 \pm 0.0004) \times 10^6$ MeV$^2$. Subsequently, by employing this $\sigma_{\,cc}$ value, along with the established parameters $\bar{M}_{1S}=2546\pm5.6~\text{MeV}$ and $\lambda = 1.479 \pm 0.015$, we use Eq. [\ref{x1}-\ref{x3}]  to extract the spin-averaged masses for the excited diquark states. The results are $\bar{M}_{2S}=3047 \pm 6$ MeV, $\bar{M}_{1P}=2932 \pm 5$ MeV, and $\bar{M}_{2P}=3252 \pm 6$ MeV.}

{As previously noted, these excited diquarks are treated as spin-singlets ($S_h = 0$) for the $1P$ and $2P$ waves and as a spin-triplet ($S_h = 1$) for the $2S$ wave. In the baryon, this diquark is connected to the light quark by a flux tube, with the system revolving about its center of mass. We extract the dynamical masses for these rotating diquarks, which have a velocity $v_{h} = 0.418 \pm 0.007$, obtaining $m_{h_{2S}} = 3354 \pm 14\ \text{MeV}$, $m_{h_{1P}} = 3227 \pm 13\ \text{MeV}$, and $m_{h_{2P}} = 3580 \pm 14\ \text{MeV}$.}

{Finally, utilizing these calculated dynamical masses for the excited diquarks, we proceed to compute the spectra for the $\Xi_{cc}$ and $\Omega_{cc}$ baryons. This computation addresses states where the light quark is orbitally or radially excited relative to the diquark in its $2S$, $1P$, or $2P$ internal wave. The calculations rely on Eqs.[\ref{eq10}, \ref{eq10.1}, \ref{eq10.2}, \ref{eq10.3}], incorporating the full set of previously fitted parameters ($m_{u/d}$, $m_s$, $\lambda$, $b_{0}$, $\sigma_0$, $v_h$, $\sigma_{\Xi_{cc}}$, and $\sigma_{\Omega_{cc}}$).}

	\begin{widetext}

	\begin{table*}
		\caption{\label{tbl1}
			Masses of $D$-Meson (in MeV).}
	
		\begin{ruledtabular}
			
			\begin{tabular}{clccccc}
				$n L (J^{P}, j) $   & MASS            &       PDG \cite{PDG2024}        & Ref.\cite{Godfrey:2015dva} & Ref. \cite{Ebert:2009ua} & Ref. \cite{Song:2015fha} \\ \hline
				$1 S (0^{-}, 1/2) $  & 1867  $\pm$   9 & 1867.25$\pm$0.035 &     1877      &1871  &1855  \\
				$1 S (1^{-}, 1/2) $  & 2009  $\pm$   8 & 2008.55$\pm$0.035 &     2041      &2010  &2020  \\
				$2 S (0^{-}, 1/2) $  & 2594	$\pm$	6    &    2549$\pm$19    &     2581      &2581  &2534  \\
				$2 S (1^{-}, 1/2) $  & 2597	$\pm$	6    &    2627$\pm$10    &     2643      &2632  & 2593 \\
				$3 S (0^{-}, 1/2) $  & 2995	$\pm$	8    &2971.8	$\pm$ 8.7                     &     3068      &3062  &2976  \\
				$3 S (1^{-}, 1/2) $  & 2995	$\pm$	8    & 3008.1	$\pm$ 4.0                 &     3110      &3096  &3015  \\
				$4 S (0^{-}, 1/2) $  & 3314	$\pm$	9    &                   &     3468      &3452  &  \\
				$4 S (1^{-}, 1/2) $  & 3314	$\pm$	9    &                   &     3497      &3482  &  \\
				$5 S (0^{-}, 1/2) $  & 3587	$\pm$	11   &                   &     3814      & 3793 &  \\
				$5 S (1^{-}, 1/2) $  & 3587	$\pm$	11   &                   &     3837      &3822  &  \\
				$6 S (0^{-}, 1/2) $  & 3830	$\pm$	12   &                   &               &  &  \\
				$6 S (1^{-}, 1/2) $  & 3830	$\pm$	12   &                   &               &  &  \\
				$1 P (0^{+}, 1/2) $  & 2362	$\pm$	7    &    2343$\pm$10    &     2399      &2406  &2365  \\
				$1 P (1^{+}, 1/2) $  & 2422	$\pm$	6    &  2422.1$\pm$0.6   &     2456      &2469  & 2426 \\
				$1 P (1^{+}, 3/2) $  & 2429	$\pm$	5    &    2412$\pm$9     &     2467      &2426  & 2431 \\
				$1 P (2^{+}, 3/2) $  & 2462	$\pm$	4    &  2461.1$\pm$0.8   &     2502      &2460  &2468  \\
				$2 P (0^{+}, 1/2) $  & 2873	$\pm$	6    &                   &     2931      &2919  &2856  \\
				$2 P (1^{+}, 1/2) $  & 2888	$\pm$	5    &                   &     2924      &3021  &2861  \\
				$2 P (1^{+}, 3/2) $  & 2871	$\pm$	5    &                   &     2961      &2932  &2877  \\
				$2 P (2^{+}, 3/2) $  & 2877	$\pm$	5    &                   &     2957      &3012  &2884 \\
				$3 P (0^{+}, 1/2) $  & 3222	$\pm$	7    &                   &     3343      &3346  &  \\
				$3 P (1^{+}, 1/2) $  & 3228	$\pm$	7    &                   &     3328      & 3461 &  \\
				$3 P (1^{+}, 3/2) $  & 3211	$\pm$	7    &                   &     3360      &3365  &  \\
				$3 P (2^{+}, 3/2) $  & 3212	$\pm$	7    & 3214	$\pm$ 29 $\pm$ 49                 &     3353      &3407  &  \\
				$4 P (0^{+}, 1/2) $  & 3510	$\pm$	9    &                   &     3697      &  &  \\
				$4 P (1^{+}, 1/2) $  & 3513	$\pm$	9    &                   &     3681      &  &  \\
				$4 P (1^{+}, 3/2) $  & 3497	$\pm$	9    &                   &     3709      &  &  \\
				$4 P (2^{+}, 3/2) $  & 3497	$\pm$	9    &                   &     3701      &  &  \\
				$1 D (1^{-}, 3/2) $  & 2739	$\pm$	5    &    2781$\pm$22    &     2817      &2788  &2762  \\
				$1 D (2^{-}, 3/2) $  & 2763	$\pm$	4    &    2747$\pm$6     &     2816      &2850  &2773  \\
				$1 D (2^{-}, 5/2) $  & 2738	$\pm$	4    &    2747$\pm$6     &     2845      &2806  &2779  \\
				$1 D (3^{-}, 5/2) $  & 2754	$\pm$	4    &  2763.1$\pm$3.2   &     2833      &2863  &2779  \\
				$2 D (1^{-}, 3/2) $  & 3122	$\pm$	5    &                   &     3231      &3228  &3131  \\
				$2 D (2^{-}, 3/2) $  & 3131	$\pm$	5    &                   &     3212      &3307  &3128  \\
				$2 D (2^{-}, 5/2) $  & 3102	$\pm$	5    &                   &     3248      &3259 &3136  \\
				$2 D (3^{-}, 5/2) $  & 3107	$\pm$	4    &                   &     3226      &3335  &3129  \\
				$3 D (1^{-}, 3/2) $  & 3427	$\pm$	6    &                   &     3588      &  &  \\
				$3 D (2^{-}, 3/2) $  & 3431	$\pm$	6    &                   &     3566      &  &  \\
				$3 D (2^{-}, 5/2) $  & 3403	$\pm$	6    &                   &     3600      &  &  \\
				$3 D (3^{-}, 5/2) $  & 3405	$\pm$	6    &                   &     3579      &  &  \\
				$1 F (2^{+}, 5/2) $  & 3013	$\pm$	4    &  3008.1	$\pm$ 4.0                &     3132      &3090  &3053  \\
				$1 F (3^{+}, 5/2) $  & 3028	$\pm$	4    &                   &     3108      &3145  &3046  \\
				$1 F (3^{+}, 7/2) $  & 2987	$\pm$	3    & 2971.8$\pm$8.7                  &     3143      &3129  &3051  \\
				$1 F (4^{+}, 7/2) $  & 2997	$\pm$	3    &3008.1$\pm$4.0                   &     3113      &3187  &3037  \\
				$2 F (2^{+}, 5/2) $  & 3338	$\pm$	4    &                   &     3490      &  &  \\
				$2 F (3^{+}, 5/2) $  & 3345	$\pm$	4    &                   &     3461      &  &  \\
				$2 F (3^{+}, 7/2) $  & 3305	$\pm$	4    &                   &     3498      &3551  &  \\
				$2 F (4^{+}, 7/2) $  & 3308	$\pm$	4    &                   &     3466      &3610  &  \\
				$1 G (3^{-}, 7/2) $  & 3244	$\pm$	4    &                   &     3397      &3352  &  \\
				$1 G (4^{-}, 7/2) $  & 3255	$\pm$	3    &                   &     3364      &3415  &  \\
				$1 G (4^{-}, 9/2) $  & 3202	$\pm$	3    &                   &     3399      &3403  &  \\
				$1 G (5^{-}, 9/2) $  & 3209	$\pm$	3    &                   &     3362      &3473  &  \\
				$1 H (4^{+}, 9/2) $  & 3449	$\pm$	3    &                   &               &  &  \\
				$1 H (5^{+}, 9/2) $  & 3456	$\pm$	3    &                   &               &  &  \\
				$1 H (5^{+}, 11/2) $ & 3394	$\pm$	3    &                   &               &  &  \\
				$1 H (6^{+}, 11/2) $ & 3399	$\pm$	3    &                   &               &  &  
				\end{tabular}
		\end{ruledtabular}
	\end{table*}  	
	
	\begin{table*}
		\caption{\label{tbl2}	Masses of $D_s$-Meson(in MeV).}
		\begin{ruledtabular}
			\begin{tabular}{cccccc}
				$n L (J^{P}, j) $   &     MASS      &       PDG \cite{PDG2024}       & Ref.\cite{Godfrey:2015dva} & Ref. \cite{Ebert:2009ua} &  Ref. \cite{Song:2015nia}  \\ \hline
				$1 S (0^{-}, 1/2) $  & 1969 $\pm$ 9  & 1968.35$\pm$0.07 &     1979      &1969  &1967  \\
				$1 S (1^{-}, 1/2) $  & 2087 $\pm$ 8  &  2112.2$\pm$0.4  &     2129      & 2111 &2115  \\
				$2 S (0^{-}, 1/2) $  & 2709	$\pm$	7  &    2591$\pm$9    &     2673      & 2688 &2646  \\
				$2 S (1^{-}, 1/2) $  & 2714	$\pm$	7  &    2714$\pm$5    &     2732      & 2731 &2704  \\
				$3 S (0^{-}, 1/2) $  & 3139	$\pm$	8  &                  &     3154      &3219  &  \\
				$3 S (1^{-}, 1/2) $  & 3139	$\pm$	8  &                  &     3193      &3242  &  \\
				$4 S (0^{-}, 1/2) $  & 3481	$\pm$	10 &                  &     3547      &3652  &  \\
				$4 S (1^{-}, 1/2) $  & 3481	$\pm$	10 &                  &     3575      &3669  &  \\
				$5 S (0^{-}, 1/2) $  & 3775	$\pm$	11 &                  &     3894      &4033  &  \\
				$5 S (1^{-}, 1/2) $  & 3775	$\pm$	11 &                  &     3912      &4048  &  \\
				$6 S (0^{-}, 1/2) $  & 4036	$\pm$	13 &                  &               &  &  \\
				$6 S (1^{-}, 1/2) $  & 4036	$\pm$	13 &                  &               &  &  \\
				$1 P (0^{+}, 1/2) $  & 2461	$\pm$	7  &  2317.8$\pm$0.5  &     2484      &2509  &2463  \\
				$1 P (1^{+}, 1/2) $  & 2524	$\pm$	6  &  2459.5$\pm$0.6  &     2549      &2574  &2531  \\
				$1 P (1^{+}, 3/2) $  & 2535	$\pm$	5  & 2535.11$\pm$0.06 &     2556      &2536  &2532  \\
				$1 P (2^{+}, 3/2) $  & 2572	$\pm$	4  &  2569.1$\pm$0.8  &     2592      &2571  &2571  \\
				$2 P (0^{+}, 1/2) $  & 3003	$\pm$	6  &                  &     3005      &3054  &2960  \\
				$2 P (1^{+}, 1/2) $  & 3019	$\pm$	6  & 3044$^{+31}_{-9}$                 &     3018      &3154  &2979  \\
				$2 P (1^{+}, 3/2) $  & 3009	$\pm$	5  &                  &     3038      &3067  &2988  \\
				$2 P (2^{+}, 3/2) $  & 3015	$\pm$	5  &                  &     3048      &3142  &3004  \\
				$3 P (0^{+}, 1/2) $  & 3377	$\pm$	8  &                  &     3412      &3513  &  \\
				$3 P (1^{+}, 1/2) $  & 3384	$\pm$	7  &                  &     3416      &3618  &  \\
				$3 P (1^{+}, 3/2) $  & 3373	$\pm$	7  &                  &     3433      &3519  &  \\
				$3 P (2^{+}, 3/2) $  & 3374	$\pm$	7  &                  &     3439      &3580  &  \\
				$4 P (0^{+}, 1/2) $  & 3687	$\pm$	9  &                  &     3764      &  &  \\
				$4 P (1^{+}, 1/2) $  & 3691	$\pm$	9  &                  &     3764      &  &  \\
				$4 P (1^{+}, 3/2) $  & 3680	$\pm$	9  &                  &     3778      &  &  \\
				$4 P (2^{+}, 3/2) $  & 3680	$\pm$	9  &                  &     3783      &  &  \\
				$1 D (1^{-}, 3/2) $  & 2857	$\pm$	5  &   2859$\pm$27    &     2899      &2913  &2865  \\
				$1 D (2^{-}, 3/2) $  & 2883	$\pm$	4  &                  &     2900      &2961  &2877  \\
				$1 D (2^{-}, 5/2) $  & 2869	$\pm$	4  &                  &     2926      &2931  &2882  \\
				$1 D (3^{-}, 5/2) $  & 2886	$\pm$	4  &    2860$\pm$7    &     2917      &2971  &2883  \\
				$2 D (1^{-}, 3/2) $  & 3268	$\pm$	5  &                  &     3306      &3383  &3244  \\
				$2 D (2^{-}, 3/2) $  & 3278	$\pm$	5  &                  &     3298      &3456  &3247  \\
				$2 D (2^{-}, 5/2) $  & 3259	$\pm$	5  &                  &     3323      &3403  &3252  \\
				$2 D (3^{-}, 5/2) $  & 3264	$\pm$	5  &                  &     3311      &3469  &3251  \\
				$3 D (1^{-}, 3/2) $  & 3596	$\pm$	7  &                  &     3658      &  &  \\
				$3 D (2^{-}, 3/2) $  & 3600	$\pm$	7  &                  &     3650      &  &  \\
				$3 D (2^{-}, 5/2) $  & 3581	$\pm$	7  &                  &     3672      &  &  \\
				$3 D (3^{-}, 5/2) $  & 3583	$\pm$	7  &                  &     3661      &  &  \\
				$1 F (2^{+}, 5/2) $  & 3148	$\pm$	4  &                  &     3208      &3230  &3159  \\
				$1 F (3^{+}, 5/2) $  & 3164	$\pm$	3  &                  &     3186      &3266  &3151  \\
				$1 F (3^{+}, 7/2) $  & 3138	$\pm$	3  &                  &     3218      &3254  &3157  \\
				$1 F (4^{+}, 7/2) $  & 3149	$\pm$	3  &                  &     3190      &3300  &3143  \\
				$2 F (2^{+}, 5/2) $  & 3498	$\pm$	5  &                  &     3562      &  &  \\
				$2 F (3^{+}, 5/2) $  & 3505	$\pm$	4  &                  &     3540      &  &  \\
				$2 F (3^{+}, 7/2) $  & 3478	$\pm$	4  &                  &     3569      &3710  &  \\
				$2 F (4^{+}, 7/2) $  & 3482	$\pm$	4  &                  &     3544      &3754  &  \\
				$1 G (3^{-}, 7/2) $  & 3394	$\pm$	3  &                  &     3469      &3508  &  \\
				$1 G (4^{-}, 7/2) $  & 3405	$\pm$	3  &                  &     3436      &3554  &  \\
				$1 G (4^{-}, 9/2) $  & 3370	$\pm$	3  &                  &     3469      &3546  &  \\
				$1 G (5^{-}, 9/2) $  & 3378	$\pm$	3  &                  &     3433      &3595  &  \\
				$1 H (4^{+}, 9/2) $  & 3613	$\pm$	3  &                  &               &  &  \\
				$1 H (5^{+}, 9/2) $  & 3620	$\pm$	3  &                  &               &  &  \\
				$1 H (5^{+}, 11/2) $ & 3578	$\pm$	3  &                  &               &  &  \\
				$1 H (6^{+}, 11/2) $ & 3583	$\pm$	3  &                  &               &  &		
				\end{tabular}
		\end{ruledtabular}
	\end{table*}  
	
\begin{table*}
	\caption{\label{tbl3} Masses of $\Xi_{cc}$-baryon (in MeV).}
	\begin{ruledtabular}
		\begin{tabular}{ccccccccccc}
			      $n_dL_dnL(J^{P},j)$       &          MASS           & PDG \cite{PDG2024} & {\cite{Bahtiyar:2020uuj} }& \cite{Ebert:2002ig} & \cite{Oudichhya:2022ssc} & \cite{Roberts:2007ni} & \cite{Yoshida:2015tia} & \cite{Giannuzzi:2009gh} & \cite{Valcarce:2008dr} &  \\ \hline
			      $1S1s(1/2^{+},1/2)$       &      3622 $\pm$ 16      &  3621.6$\pm$ 0.4   & {3615 $\pm$ 33}                      &        3620         &           3581           &         3676          &          3685          &          3547           &          3579          &  \\
			      $1S1s(3/2^{+},1/2)$       &      3727 $\pm$ 16      &                    & {3703 $\pm$ 33}                     &        3727         &           3726           &         3753          &          3754          &          3719           &          3656          &  \\
			      $1S2s(1/2^{+},1/2)$       &      4222 $\pm$ 11      &                    &                                &                     &           3925           &         4029          &          4079          &          4183           &          3910          &  \\
			      $1S2s(3/2^{+},1/2)$       &      4225 $\pm$ 11      &                    &                                &                     &           3988           &         4042          &          4114          &          4282           &          4027          &  \\
			      $1S1p(1/2^{-},1/2)$       &      4027 $\pm$ 13      &                    & {3930 $\pm$ 20}                       &        4053         &                          &         3910          &          3947          &                         &          3880          &  \\
			      $1S1p(1/2^{-},3/2)$       &      4064 $\pm$ 12      &                    & {3971 $\pm$ 22 }                      &        4136         &                          &       4074                &    4135                    &                         &                        &  \\
			      $1S1p(3/2^{-},1/2)$       &      4073 $\pm$ 12      &                    & {4009 $\pm$ 31}                       &        4101         &           3926           &         3921          &          3949          &                         &                        &  \\
			      $1S1p(3/2^{-},3/2)$       &      4077 $\pm$ 11      &                    &                                &        4196         &                          &       4078                &    4137                    &                         &                        &  \\
			      $1S1p(5/2^{-},3/2)$       &      4099 $\pm$ 11      &                    &                                &        4155         &           4074           &         4092          &          4163          &                         &                        &  \\
			{${1P1s(1/2^{-}, 1/2)}$} & {4130 $\pm$ 24}  &                    &{ 4246 $\pm$ 193 }                  &        3838         &                          &                       &   4149                     &                         &                        &  \\
			 ${1P1s(3/2^{-}, 1/2)}$  & {4221 $\pm$  24} &                    &                                &        3959         &                          &                       &     4159                   &                         &                        &  \\
			 ${2S1s(1/2^{+}, 1/2)}$  & {4281 $\pm$ 26}  &                    &                                &        3910         &                          &                       &   4159                     &                         &                        &  \\
			 ${2S1s(3/2^{+}, 1/2)}$  & {4369 $\pm$ 25}  &                    &                                & 4027                &                          &                       &  4131                      &                         &                        &  \\
			    $1S2 p (1/2^{-}, 1/2) $     &      4493 $\pm$ 10      &                    &                                &                     &                          &                   &                    &                         &          4018          &  \\
			    $1S2 p (1/2^{-}, 3/2) $     &      4483 $\pm$ 10      &                    &                                &                     &                          &                       &                        &                         &                        &  \\
			    $1S2 p (3/2^{-}, 1/2) $     &      4505 $\pm$ 10      &                    &                                &                     &           4144           &                   &          4137          &                         &                        &  \\
			    $1S2 p (3/2^{-}, 3/2) $     &      4486 $\pm$ 9       &                    &                                &                     &                          &                       &                        &                         &                        &  \\
			    $1S2 p (5/2^{-}, 3/2) $     &      4490 $\pm$ 9       &                    &                                &                     &           4183           &                       &          4488          &                         &                        &  \\
			      $1S1d(1/2^{+},3/2)$       &      4362 $\pm$ 10      &                    &                                &                     &                          &                       &                        &                         &                        &  \\
			      $1S1d(3/2^{+},3/2)$       &      4372 $\pm$ 10      &                    &                                &                     &                          &                   &                        &                         &                        &  \\
			      $1S1d(3/2^{+},5/2)$       &      4353 $\pm$ 9       &                    &                                &                     &                          &                       &                        &                         &                        &  \\
			      $1S1d(5/2^{+},3/2)$       &      4389 $\pm$ 10      &                    &                                &                     &           4243           &         4047          &          4115          &                         &                        &  \\
			      $1S1d(5/2^{+},5/2)$       &      4360 $\pm$ 9       &                    &                                &                     &                          & 4091                      &   4164                     &                         &                        &  \\
			      $1S1d(7/2^{+},5/2)$       &      4370 $\pm$ 9       &                    &                                &                     &           4394           &         4097          &                        &                         &                        &  \\
			 ${2P1s(1/2^{-}, 1/2)}$  & {4550 $\pm$ 27}  &                    &                                &        4085         &                          &                       &                        &                         &                        &  \\
			 ${2P1s(3/2^{-}, 1/2)}$  & { 4633 $\pm$ 27} &                    &                                &        4197         &                          &                       &                        &                         &                        &  \\
			    $1S2 d (1/2^{+}, 3/2) $     &      4732 $\pm$ 9       &                    &                                &                     &                          &                       &                        &                         &                        &  \\
			    $1S2 d (3/2^{+}, 3/2) $     &      4736 $\pm$ 9       &                    &                                &                     &                          &                   &                        &                         &                        &  \\
			    $1S2 d (3/2^{+}, 5/2) $     &      4707 $\pm$ 8       &                    &                                &                     &                          &                       &                        &                         &                        &  \\
			    $1S2 d (5/2^{+}, 3/2) $     &      4743 $\pm$ 8       &                    &                                &                     &           4299           &                   &                    &                         &                        &  \\
			    $1S2 d (5/2^{+}, 5/2) $     &      4710 $\pm$ 8       &                    &                                &                     &                          &                       &                        &                         &                        &  \\
			    $1S2 d (7/2^{+}, 5/2) $     &      4714 $\pm$ 8       &                    &                                &                     &                          &                   &                        &                         &                        &  \\
			    $1S1 f (3/2^{-}, 5/2) $     &      4626 $\pm$ 9       &                    &                                &                     &                          &                       &                        &                         &                        &  \\
			    $1S1 f (5/2^{-}, 5/2) $     &      4633 $\pm$ 8       &                    &                                &                     &                          &                       &   4534                     &                         &                        &  \\
			    $1S1 f (5/2^{-}, 7/2) $     &      4593 $\pm$ 8       &                    &                                &                     &                          &                       &                        &                         &                        &  \\
			    $1S1 f (7/2^{-}, 5/2) $     &      4644 $\pm$ 8       &                    &                                &                     &           4538           &                       &                        &                         &                        &  \\
			    $1S1 f (7/2^{-}, 7/2) $     &      4598 $\pm$ 8       &                    &                                &                     &                          &                       &                        &                         &                        &  \\
			    $1S1 f (9/2^{-}, 7/2) $     &      4605 $\pm$ 8       &                    &                                &                     &           4692           &                       &                        &                         &                        &  \\
			    $1S1 g (5/2^{+}, 7/2) $     &      4853 $\pm$ 8       &                    &                                &                     &                          &                       &                        &                         &                        &  \\
			    $1S1 g (7/2^{+}, 7/2) $     &      4859 $\pm$ 8       &                    &                                &                     &                          &                       &                        &                         &                        &  \\
			    $1S1 g (7/2^{+}, 9/2) $     &      4804 $\pm$ 8       &                    &                                &                     &                          &                       &                        &                         &                        &  \\
			    $1S1 g (9/2^{+}, 7/2) $     &      4866 $\pm$ 7       &                    &                                &                     &           4815           &                       &                        &                         &                        &  \\
			    $1S1 g (9/2^{+}, 9/2) $     &      4808 $\pm$ 7       &                    &                                &                     &                          &                       &                        &                         &                        &  \\
			   $1S1 g (11/2^{+}, 9/2) $     &      4814 $\pm$ 7       &                    &                                &                     &           4973           &                       &                        &                         &                        &
		\end{tabular}
	\end{ruledtabular}
\end{table*}

\begin{table*}
	\caption{\label{tbl4}	Masses of $\Omega_{cc}$-baryon	(in MeV).}	
	
	\begin{ruledtabular}
		
		\begin{tabular}{cccclllllccc}
			  $n_dL_dnL(J^{P},j)$    &       MASS        & {\cite{Bahtiyar:2020uuj}}& \cite{Ebert:2002ig} & \cite{Oudichhya:2022ssc} & \cite{Roberts:2007ni} & \cite{Yoshida:2015tia} & \cite{Valcarce:2008dr} & \cite{Giannuzzi:2009gh} &  &  \\ \hline
 			  $1S1s(1/2^{+},1/2)$    & 3718   $\pm$   16 & {3733 $\pm$ 13}&        3778         & 3719                     & 3815                  & 3832                   & 3697                   & 3648                    &  &  \\
			  $1S1s(3/2^{+},1/2)$    & 3805   $\pm$   16 & {3793 $\pm$ 30}&        3872         & 3847                     & 3876                  & 3883                   & 3769                   & 3770                    &  &  \\
			  $1S2s(1/2^{+},1/2)$    &   4337	$\pm$	12   & &                     & 4044                     & 4180                  & 4227                   & 4112                   & 4268                    &  &  \\
			  $1S2s(3/2^{+},1/2)$    &   4340	$\pm$	11   & &                     & 4096                     & 4188                  & 4263                   & 4160                   & 4334                    &  &  \\
			  $1S1p(1/2^{-},1/2)$    &   4129	$\pm$	13   &{4041 $\pm$ 15} &        4208         &                          & 4046                  & 4086                   & 4009                   &                         &  &  \\
			  $1S1p(1/2^{-},3/2)$    &   4170	$\pm$	12   &{4063 $\pm$ 13} &        4271         &                          &   4135                    &  4199                      &                        &                         &  &  \\
			  $1S1p(3/2^{-},1/2)$    &   4176	$\pm$	12   &{4115 $\pm$ 70} &        4252         & 3947                     & 4052                  & 4086                   &                        &                         &  &  \\
			  $1S1p(3/2^{-},3/2)$    &   4184	$\pm$	11   & &        4325         &                 & 4140 &4201  \\
			  $1S1p(5/2^{-},3/2)$    &   4207	$\pm$	11   & &        4303         & 4153                     & 4152                  & 4220                   &                        &                         &  &  \\
			  ${1P1s(1/2^{-}, 1/2)}$   &   {4225 $\pm$ 24}                & {4395 $\pm$ 41}&   4002                  &                          &                       &  4210                      &                        &                         &  &  &  \\
			  ${1P1s(3/2^{-}, 1/2)}$   &   {4300 $\pm$ 24}                & &    4102                 &                          &                       &                        &                        &                         &  &  &  \\
			  ${2S1s(1/2^{+}, 1/2)}$   &  {4376 $\pm$ 26}                 & &  4075                   &                          &                       &     4295                   &                        &                         &  &  &  \\
			  ${2S1s(3/2^{+}, 1/2)}$   &  {4449 $\pm$ 25}                 & &   4174                  &                          &                       &      4265                  &                        &                         &  &  &  \\
			$1S2 p (1/2^{-}, 1/2) $  &   4620	$\pm$	10   & &                     & 4135                     &                   &                    & 4101                   &                         &  &  \\
			$1S2 p (1/2^{-}, 3/2) $  &   4617	$\pm$	10   & &                     &                          &                       &                        &                        &                         &  &  \\
			$1S2 p (3/2^{-}, 1/2) $  &   4632	$\pm$	10   & &                     & 4259                     &                   &                    &                        &                         &  &  \\
			$1S2 p (3/2^{-}, 3/2) $  &   4620	$\pm$	10   & &                     &                          &                       &                        &                        &                         &  &  \\
			$1S2 p (5/2^{-}, 3/2) $  &   4624	$\pm$	10   & &                     & 4247                     &                       & 4555                   &                        &                         &  &  \\
			$1S1 d (1/2^{+}, 3/2) $  &   4479	$\pm$	10   & &                     &                          &                       &                        &                        &                         &  &  \\
			$1S1 d (3/2^{+}, 3/2) $  &   4489	$\pm$	10   & &                     &                          &                       &                        &                        &                         &  &  \\
			$1S1 d (3/2^{+}, 5/2) $  &   4481	$\pm$	10   & &                     &                          &                       &                        &                        &                         &  &  \\
			$1S1 d (5/2^{+}, 3/2) $  &   4506	$\pm$	10   & &                     & 4162                     & 4202                  &   4264                     &                        &                         &  &  \\
			$1S1 d (5/2^{+}, 5/2) $  &   4488	$\pm$	9    & &                     &                          &  4232                     &   4299                     &                        &                         &  &  \\
			$1S1 d (7/2^{+}, 5/2) $  &   4499	$\pm$	9    & &                     & 4438                     & 4230                  &                        &                        &                         &  &  \\
			  ${2P1s(1/2^{-}, 1/2)}$   &  {4644 $\pm$ 27}                 & &  4251                   &                          &                       &                        &                        &                         &  &  &  \\
			  ${2P1s(3/2^{-}, 1/2)}$   &  {4713 $\pm$ 27}                 & &    4345                 &                          &                       &                        &                        &                         &  &  &  \\
			$1S2 d (1/2^{+}, 3/2) $  &   4873	$\pm$	9    & &                     &                          &                       &                        &                        &                         &  &  \\
			$1S2 d (3/2^{+}, 3/2) $  &   4877	$\pm$	9    & &                     &                          &                       &                        &                        &                         &  &  \\
			$1S2 d (3/2^{+}, 5/2) $  &   4858	$\pm$	9    & &                     &                          &                       &                        &                        &                         &  &  \\
			$1S2 d (5/2^{+}, 3/2) $  &   4883	$\pm$	9    & &                     & 4407                     &                   &                        &                        &                         &  &  \\
			$1S2 d (5/2^{+}, 5/2) $  &   4861	$\pm$	9    & &                     &                          &                       &                        &                        &                         &  &  \\
			$1S2 d (7/2^{+}, 5/2) $  &   4865	$\pm$	8    & &                     &                          &                   &                        &                        &                         &  &  \\
			$1S1 f (3/2^{-}, 5/2) $  &   4757	$\pm$	9    & &                     &                          &                       &                        &                        &                         &  &  \\
			$1S1 f (5/2^{-}, 5/2) $  &   4764	$\pm$	8    & &                     &                          &                       &                        &                        &                         &  &  \\
			$1S1 f (5/2^{-}, 7/2) $  &   4739	$\pm$	8    & &                     &                          &                       &   4600                     &                        &                         &  &  \\
			$1S1 f (7/2^{-}, 5/2) $  &   4775	$\pm$	8    & &                     & 4367                     &                       &                        &                        &                         &  &  \\
			$1S1 f (7/2^{-}, 7/2) $  &   4745	$\pm$	8    & &                     &                          &                       &                        &                        &                         &  &  \\
			$1S1 f (9/2^{-}, 7/2) $  &   4752	$\pm$	8    & &                     & 4706                     &                       &                        &                        &                         &  &  \\
			$1S1 g (5/2^{+}, 7/2) $  &   4997	$\pm$	8    & &                     &                          &                       &                        &                        &                         &  &  \\
			$1S1 g (7/2^{+}, 7/2) $  &   5003	$\pm$	8    & &                     &                          &                       &                        &                        &                         &  &  \\
			$1S1 g (7/2^{+}, 9/2) $  &   4966	$\pm$	8    & &                     &                          &                       &                        &                        &                         &  &  \\
			$1S1 g (9/2^{+}, 7/2) $  &   5010	$\pm$	7    & &                     & 4562                     &                       &                        &                        &                         &  &  \\
			$1S1 g (9/2^{+}, 9/2) $  &   4971	$\pm$	7    & &                     &                          &                       &                        &                        &                         &  &  \\
			$1S1 g (11/2^{+}, 9/2) $ &   4976	$\pm$	7    & &                     & 4960                     &                       &                        &                        &                         &\\
		\end{tabular}
	\end{ruledtabular}
\end{table*}

\begin{table*}
	\caption{\label{tbld1} 	{Partial strong decay widths of the $1P$, $1D$, and $2S$ states of the $D$ and $D_s$ meson families in terms of the relevant strong couplings. The total width is the sum over all listed channels. The branching fraction (in percent) refers to the ratio of each partial width to the corresponding total width.}	}	
	\begin{ruledtabular}	
		\begin{tabular}{clcclcc}
			States& \multicolumn{3}{c}{$D$ meson} & \multicolumn{3}{c}{$D_s$ meson} \\
			\cline{2-4}\cline{5-7}
			\multicolumn{1}{c}{$nL(J^{P},j)$} &
			\multicolumn{1}{c}{Channel} &
			\multicolumn{1}{c}{Width} &
			\multicolumn{1}{c}{Branching fraction} &
			\multicolumn{1}{c}{Channel} &
			\multicolumn{1}{c}{Width} &
			\multicolumn{1}{c}{Branching fraction} \\
			\hline
			\multirow{9}{*}{$1P(2^{+},3/2)$} 
			& $D^{*+}\pi^-$    & 55.35 $g_{TH}^2$  & 19.96
			& $D_{s}^{*+}K^0$      &3.84 $g_{TH}^2$  &3.29 \\ 
			& $D^{*0}\pi^0$    & 29.18 $g_{TH}^2$ & 10.52
			& $D_{s}^{*0}K^+$      &5.01  $g_{TH}^2$ &4.29\\ 
			& $D^{*0}\eta$     &-  & -
			& $D_{s}^{*+}\eta$     &-  & -\\ 
			& $D_{s}^{*+}K^-$  &-  &-
			& $D_{s}^{*+}\pi^0$    &0.0033 $g_{TH}^2$   & 0.0028\\ 
			& $D^{+}\pi^-$    &126.46 $g_{TH}^2$   & 45.61
			& $D_{s}^{+}K^0$      &50.74 $g_{TH}^2$  &43.42\\ 
			& $D^{0}\pi^0$    &66.01 $g_{TH}^2$   & 23.80
			& $D_{s}^{0}K^+$      &55.90  $g_{TH}^2$ &47.83  \\ 
			& $D^{0}\eta$     &0.24 $g_{TH}^2$  &0.08 
			& $D_{s}^{+}\eta$     &1.37 $g_{TH}^2$  &1.17\\ 
			& $D_{s}^{+}K^-$  &-  &-
			& $D_{s}^{+}\pi^0$    &0.0072 $g_{TH}^2$  & 0.0062\\ 
			& \textbf{Total}   & 277.25 $g_{TH}^2$ &
			& \textbf{Total}   & 116.87  $g_{TH}^2$ & \\
			\hline
			\multirow{9}{*}{$1D(1^{-},3/2)$} 
			& $D^{*+}\pi^-$    &295.29 $g_{XH}^2$  & 10.68
			& $D_{s}^{*+}K^0$      &336.82 $g_{XH}^2$  & 8.19\\ 
			& $D^{*0}\pi^0$    &150.86 $g_{XH}^2$  & 5.46
			& $D_{s}^{*0}K^+$      &347.61 $g_{XH}^2$  &8.45 \\ 
			& $D^{*0}\eta$     & 17.72 $g_{XH}^2$ & 0.64
			& $D_{s}^{*+}\eta$     & 88.58 $g_{XH}^2$  & 2.15\\ 
			& $D_{s}^{*+}K^-$  & 49.38 $g_{XH}^2$ & 1.79
			& $D_{s}^{*+}\pi^0$    &0.017 $g_{XH}^2$  & 0.00\\ 
			& $D^{+}\pi^-$    &1160.27 $g_{XH}^2$  & 41.98
			& $D_{s}^{+}K^0$      &1399.88 $g_{XH}^2$  & 34.04\\ 
			& $D^{0}\pi^0$    &592.77 $g_{XH}^2$  & 21.45
			& $D_{s}^{0}K^+$      & 1438.31 $g_{XH}^2$  &34.97 \\ 
			& $D^{0}\eta$     &108.72 $g_{XH}^2$  & 3.93
			& $D_{s}^{+}\eta$     & 501.52 $g_{XH}^2$ & 12.19\\ 
			& $D_{s}^{+}K^-$  &388.59 $g_{XH}^2$  &14.06
			& $D_{s}^{+}\pi^0$    &0.066 $g_{XH}^2$  & 0.00\\ 
			& \textbf{Total}   & 2763.63 $g_{XH}^2$ &
			& \textbf{Total}   &4112.81 $g_{XH}^2$  & \\
			\hline
			\multirow{5}{*}{$1D(2^{-},3/2)$} 
			& $D^{*+}\pi^-$    &1017.04 $g_{XH}^2$  & 56.30
			& $D_{s}^{*+}K^0$      &1192.26 $g_{XH}^2$  &43.21 \\ 
			& $D^{*0}\pi^0$    & 519.09 $g_{XH}^2$ & 28.74
			& $D_{s}^{*0}K^+$      &1227.68 $g_{XH}^2$  & 44.49\\ 
			& $D^{*0}\eta$     &67.80 $g_{XH}^2$  & 3.75
			& $D_{s}^{*+}\eta$     &339.59 $g_{XH}^2$  & 12.31\\ 
			& $D_{s}^{*+}K^-$  &202.38 $g_{XH}^2$   &11.20
			& $D_{s}^{*+}\pi^0$    &0.060 $g_{XH}^2$   &0.00 \\  
			& \textbf{Total}   &1806.32 $g_{XH}^2$  &
			& \textbf{Total}   &2759.50 $g_{XH}^2$  & \\
			\hline
			\multirow{5}{*}{$1D(2^{-},5/2)$} 
			& $D^{*+}\pi^-$    &128.47 $g_{YH}^2$  & 64.65
			& $D_{s}^{*+}K^0$      &89.71 $g_{YH}^2$  & 45.89\\ 
			& $D^{*0}\pi^0$    &66.58 $g_{YH}^2$  & 33.51
			& $D_{s}^{*0}K^+$      &95.54 $g_{YH}^2$   & 48.87\\ 
			& $D^{*0}\eta$     &1.33 $g_{YH}^2$   & 0.67
			& $D_{s}^{*+}\eta$     & 10.22 $g_{YH}^2$ & 5.23\\ 
			& $D_{s}^{*+}K^-$  &2.31 $g_{YH}^2$  &1.16
			& $D_{s}^{*+}\pi^0$    & 0.009 $g_{YH}^2$  & 0.00\\  
			& \textbf{Total}   &198.71 $g_{YH}^2$  &
			& \textbf{Total}   & 195.49 $g_{YH}^2$ & \\
			\hline
			\multirow{9}{*}{$1D(3^{-},5/2)$} 
			& $D^{*+}\pi^-$    &84.01 $g_{YH}^2$  & 20.80
			& $D_{s}^{*+}K^0$      &61.49 $g_{YH}^2$  &13.02 \\ 
			& $D^{*0}\pi^0$    &43.49 $g_{YH}^2$  & 10.77
			& $D_{s}^{*0}K^+$      &65.28 $g_{YH}^2$  &13.83 \\ 
			& $D^{*0}\eta$     &1.05 $g_{YH}^2$   & 0.26
			& $D_{s}^{*+}\eta$     &7.80 $g_{YH}^2$   &1.65 \\ 
			& $D_{s}^{*+}K^-$  &1.99 $g_{YH}^2$  &0.49
			& $D_{s}^{*+}\pi^0$    &0.0059 $g_{YH}^2$  &0.00 \\ 
			& $D^{+}\pi^-$    & 165.855 $g_{YH}^2$ & 41.07
			& $D_{s}^{+}K^0$      &148.611 $g_{YH}^2$   &31.48 \\ 
			& $D^{0}\pi^0$    & 85.80 $g_{YH}^2$ & 21.25
			& $D_{s}^{0}K^+$      &156.28  $g_{YH}^2$  &33.10 \\ 
			& $D^{0}\eta$     & 5.56  $g_{YH}^2$ & 1.38
			& $D_{s}^{+}\eta$     & 32.62 $g_{YH}^2$ & 6.91\\ 
			& $D_{s}^{+}K^-$  &16.05 $g_{YH}^2$  &3.97
			& $D_{s}^{+}\pi^0$    & 0.011 $g_{YH}^2$ &0.00 \\ 
			& \textbf{Total}   &403.83 $g_{YH}^2$  &
			& \textbf{Total}   &472.12  $g_{YH}^2$ & \\
			\hline
			\multirow{5}{*}{$2S(0^{-},1/2)$} 
			& $D^{*+}\pi^-$    &927.69 $\tilde g_{HH}^2$  & 65.91
			& $D_{s}^{*+}K^0$      &536.04 $\tilde g_{HH}^2$  &46.60 \\ 
			& $D^{*0}\pi^0$    &472.002 $\tilde g_{HH}^2$  & 33.54
			& $D_{s}^{*0}K^+$      &561.52 $\tilde g_{HH}^2$   &48.82 \\ 
			& $D^{*0}\eta$     &7.78 $\tilde g_{HH}^2$  & 0.55
			& $D_{s}^{*+}\eta$     & 52.66 $\tilde g_{HH}^2$  &4.58 \\ 
			& $D_{s}^{*+}K^-$  &-  &-
			& $D_{s}^{*+}\pi^0$    &0.052 $\tilde g_{HH}^2$  &0.00 \\  
			& \textbf{Total}   &1407.48 $\tilde g_{HH}^2$  &
			& \textbf{Total}   &1150.27 $\tilde g_{HH}^2$  & \\
			\hline
			\multirow{9}{*}{$2S(1^{-},1/2)$} 
			& $D^{*+}\pi^-$    & 625.03 $\tilde g_{HH}^2$ & 33.69
			& $D_{s}^{*+}K^0$      &370.40 $\tilde g_{HH}^2$   &21.22 \\ 
			& $D^{*0}\pi^0$    &318.94 $\tilde g_{HH}^2$  & 17.19
			& $D_{s}^{*0}K^+$      &388.71 $\tilde g_{HH}^2$  &22.27 \\ 
			& $D^{*0}\eta$     &5.80 $\tilde g_{HH}^2$  & 0.31
			& $D_{s}^{*+}\eta$     &40.13 $\tilde g_{HH}^2$    &2.30 \\ 
			& $D_{s}^{*+}K^-$  & - &-
			& $D_{s}^{*+}\pi^0$    &0.035 $\tilde g_{HH}^2$   &0.00 \\
			& $D^{+}\pi^-$    &520.67 $\tilde g_{HH}^2$  & 28.06
			& $D_{s}^{+}K^0$      &405.05  $\tilde g_{HH}^2$  &23.20 \\ 
			& $D^{0}\pi^0$    &265.06 $\tilde g_{HH}^2$  & 14.29
			& $D_{s}^{0}K^+$      &418.31 $\tilde g_{HH}^2$   & 23.96\\ 
			& $D^{0}\eta$     &26.59  $\tilde g_{HH}^2$ & 1.43
			& $D_{s}^{+}\eta$     & 123.12 $\tilde g_{HH}^2$  &7.05 \\ 
			& $D_{s}^{+}K^-$  &93.31 $\tilde g_{HH}^2$  &5.03
			& $D_{s}^{+}\pi^0$    & 0.028 $\tilde g_{HH}^2$  &0.00 \\  
			& \textbf{Total}   &1855.40 $\tilde g_{HH}^2$ &
			& \textbf{Total}   &1745.79 $\tilde g_{HH}^2$  & \\
			
		\end{tabular}
	\end{ruledtabular}
\end{table*}

\begin{table*}
	\caption{\label{tbld2} {The same as Table \ref{tbld1}, but for $3S$ and $2P$ states of the $D$ and $D_s$ meson families}}	
	\begin{ruledtabular}	
		\begin{tabular}{clcclcc}
			States& \multicolumn{3}{c}{$D$ meson} & \multicolumn{3}{c}{$D_s$ meson} \\
			\cline{2-4}\cline{5-7}
			\multicolumn{1}{c}{$nL(J^{P},j)$} &
			\multicolumn{1}{c}{Channel} &
			\multicolumn{1}{c}{Width} &
			\multicolumn{1}{c}{Branching fraction} &
			\multicolumn{1}{c}{Channel} &
			\multicolumn{1}{c}{Width} &
			\multicolumn{1}{c}{Branching fraction} \\
			\hline
			\multirow{5}{*}{$3S(0^{-},1/2)$} 
			& $D^{*+}\pi^-$    &3412.14 $\tilde{\tilde g}_{HH}^2$  & 48.18
			& $D_{s}^{*+}K^0$      &3412.33 $\tilde{\tilde g}_{HH}^2$  &40.26 \\ 
			& $D^{*0}\pi^0$    &1716.02 $\tilde{\tilde g}_{HH}^2$  & 24.23
			& $D_{s}^{*0}K^+$      &3447.71 $\tilde{\tilde g}_{HH}^2$  &40.68 \\ 
			& $D^{*0}\eta$     &333.43  $\tilde{\tilde g}_{HH}^2$ & 4.71
			& $D_{s}^{*+}\eta$     & 1614.95 $\tilde{\tilde g}_{HH}^2$  & 19.06\\ 
			& $D_{s}^{*+}K^-$  &1620.80 $\tilde{\tilde g}_{HH}^2$ &22.88
			& $D_{s}^{*+}\pi^0$    &0.19$\tilde{\tilde g}_{HH}^2$  & 0.00\\  
			& \textbf{Total}   & 7082.40  $\tilde{\tilde g}_{HH}^2$ &
			& \textbf{Total}   &8475.18  $\tilde{\tilde g}_{HH}^2$  & \\
			\hline
			\multirow{9}{*}{$3S(1^{-},1/2)$} 
			& $D^{*+}\pi^-$    &2267.78 $\tilde{\tilde g}_{HH}^2$  & 28.67
			& $D_{s}^{*+}K^0$      &2267.90 $\tilde{\tilde g}_{HH}^2$  & 24.13\\ 
			& $D^{*0}\pi^0$    &1144.01 $\tilde{\tilde g}_{HH}^2$  & 14.46
			& $D_{s}^{*0}K^+$      &2298.47 $\tilde{\tilde g}_{HH}^2$  &24.46 \\ 
			& $D^{*0}\eta$     &222.29  $\tilde{\tilde g}_{HH}^2$ & 2.81
			& $D_{s}^{*+}\eta$     &1076.63  $\tilde{\tilde g}_{HH}^2$  & 11.46\\ 
			& $D_{s}^{*+}K^-$  &1080.54  $\tilde{\tilde g}_{HH}^2$ &13.66
			& $D_{s}^{*+}\pi^0$    &0.13$\tilde{\tilde g}_{HH}^2$  &0.00 \\ 
			& $D^{+}\pi^-$    &1455.61 $\tilde{\tilde g}_{HH}^2$  & 18.40
			& $D_{s}^{+}K^0$      &1485.37 $\tilde{\tilde g}_{HH}^2$  &15.80 \\ 
			& $D^{0}\pi^0$    &734.23 $\tilde{\tilde g}_{HH}^2$  & 9.28
			& $D_{s}^{0}K^+$      &1503.52  $\tilde{\tilde g}_{HH}^2$  & 16.00\\ 
			& $D^{0}\eta$     &164.34$\tilde{\tilde g}_{HH}^2$  & 2.08
			& $D_{s}^{+}\eta$     & 766.51 $\tilde{\tilde g}_{HH}^2$ & 8.16\\ 
			& $D_{s}^{+}K^-$  &840.84 $\tilde{\tilde g}_{HH}^2$  &10.63
			& $D_{s}^{+}\pi^0$    &0.083 $\tilde{\tilde g}_{HH}^2$  & 0.00\\  
			& \textbf{Total}   & 7909.63  $\tilde{\tilde g}_{HH}^2$ &
			& \textbf{Total}   & 9398.62  $\tilde{\tilde g}_{HH}^2$  & \\
			\hline
			\multirow{5}{*}{$2P(0^{+},1/2)$} 
			& $D^{+}\pi^-$    &3474.71  $ \tilde g_{SH}^2$ & 41.11
			& $D_{s}^{+}K^0$      &4448.32  $ \tilde g_{SH}^2$  &39.11 \\ 
			& $D^{0}\pi^0$    &1752.45  $ \tilde g_{SH}^2$  & 20.73
			& $D_{s}^{0}K^+$      &4485.08  $ \tilde g_{SH}^2$  &39.43 \\ 
			& $D^{0}\eta$     &551.23  $ \tilde g_{SH}^2$  & 6.52
			& $D_{s}^{+}\eta$     &2441.16  $ \tilde g_{SH}^2$  & 21.46\\ 
			& $D_{s}^{+}K^-$  &2654.25  $ \tilde g_{SH}^2$  &31.40
			& $D_{s}^{+}\pi^0$    &0.19  $ \tilde g_{SH}^2$  &0.00 \\ 
			& \textbf{Total}   &8432.64  $ \tilde g_{SH}^2$  &
			& \textbf{Total}   &11374.75  $ \tilde g_{SH}^2$  & \\
			\hline
			\multirow{5}{*}{$2P(1^{+},1/2)$} 
			& $D^{*+}\pi^-$    &2692.27  $ \tilde g_{SH}^2$  & 42.74
			& $D_{s}^{*+}K^0$      &3541.50  $ \tilde g_{SH}^2$  &39.59 \\ 
			& $D^{*0}\pi^0$    &1357.21  $ \tilde g_{SH}^2$   & 21.55
			& $D_{s}^{*0}K^+$      &3571.23  $ \tilde g_{SH}^2$  & 39.92\\ 
			& $D^{*0}\eta$     & 404.35  $ \tilde g_{SH}^2$ & 6.42
			& $D_{s}^{*+}\eta$     &1832.78  $ \tilde g_{SH}^2$  & 20.49\\ 
			& $D_{s}^{*+}K^-$  &1845.12  $ \tilde g_{SH}^2$  &29.29
			& $D_{s}^{*+}\pi^0$    &0.15  $ \tilde g_{SH}^2$  &0.00 \\ 
			& \textbf{Total}   &6298.96  $ \tilde g_{SH}^2$  &
			& \textbf{Total}   &8945.66  $ \tilde g_{SH}^2$  & \\
			\hline
			\multirow{5}{*}{$2P(1^{+},3/2)$} 
			& $D^{*+}\pi^-$    &1721.78 $g_{TH}^{2}$  & 57.83
			& $D_{s}^{*+}K^0$      &1608.32    $ \tilde g_{TH}^2$  &36.89 \\ 
			& $D^{*0}\pi^0$    &875.21    $ \tilde g_{TH}^2$  & 29.40
			& $D_{s}^{*0}K^+$      &1653.47    $ \tilde g_{TH}^2$  &37.93 \\ 
			& $D^{*0}\eta$     &83.31    $ \tilde g_{TH}^2$  & 2.80
			& $D_{s}^{*+}\eta$     &1097.90    $ \tilde g_{TH}^2$   &25.18 \\ 
			& $D_{s}^{*+}K^-$  &296.97    $ \tilde g_{TH}^2$  &9.97
			& $D_{s}^{*+}\pi^0$    &0.047    $ \tilde g_{TH}^2$  &0.00 \\ 
			& \textbf{Total}   &2977.27   $ \tilde g_{TH}^2$  &
			& \textbf{Total}   &4359.74   $ \tilde g_{TH}^2$  & \\
			\hline
			\multirow{9}{*}{$2P(2^{+},3/2)$} 
			& $D^{*+}\pi^-$    &1060.50    $ \tilde g_{TH}^2$  & 25.56
			& $D_{s}^{*+}K^0$      &995.14    $ \tilde g_{TH}^2$  &18.84 \\ 
			& $D^{*0}\pi^0$    &540.62    $ \tilde g_{TH}^2$  & 13.03
			& $D_{s}^{*0}K^+$      &1025.79    $ \tilde g_{TH}^2$  &19.42 \\ 
			& $D^{*0}\eta$     &52.62    $ \tilde g_{TH}^2$  & 1.27
			& $D_{s}^{*+}\eta$     &300.13    $ \tilde g_{TH}^2$  & 5.68\\ 
			& $D_{s}^{*+}K^-$  &189.25    $ \tilde g_{TH}^2$  &4.56
			& $D_{s}^{*+}\pi^0$    &0.068    $ \tilde g_{TH}^2$  &0.00 \\ 
			& $D^{+}\pi^-$    &1226.44    $ \tilde g_{TH}^2$  &29.56 
			& $D_{s}^{+}K^0$      &1233.50    $ \tilde g_{TH}^2$  &23.35 \\ 
			& $D^{0}\pi^0$    &625.23    $ \tilde g_{TH}^2$  & 15.07
			& $D_{s}^{0}K^+$      &1267.44    $ \tilde g_{TH}^2$  &24.00 \\ 
			& $D^{0}\eta$     &88.77    $ \tilde g_{TH}^2$   & 2.14
			& $D_{s}^{+}\eta$     &459.72    $ \tilde g_{TH}^2$  &8.70 \\ 
			& $D_{s}^{+}K^-$  &365.07    $ \tilde g_{TH}^2$  & 8.80
			& $D_{s}^{+}\pi^0$    &0.075    $ \tilde g_{TH}^2$  & 0.00\\ 
			& \textbf{Total}   &4148.50   $ \tilde g_{TH}^2$  &
			& \textbf{Total}   &5281.87   $ \tilde g_{TH}^2$  & \\
			\end{tabular}
	\end{ruledtabular}
\end{table*}

\begin{table*}
	\caption{\label{tbld3} {The same as Table \ref{tbld1}, but for $2F$ states of the $D$ and $D_s$ meson families}}	
	\begin{ruledtabular}	
		\begin{tabular}{clcclcc}
			States& \multicolumn{3}{c}{$D$ meson} & \multicolumn{3}{c}{$D_s$ meson} \\
			\cline{2-4}\cline{5-7}
			\multicolumn{1}{c}{$nL(J^{P},j)$} &
			\multicolumn{1}{c}{Channel} &
			\multicolumn{1}{c}{Width} &
			\multicolumn{1}{c}{Branching fraction} &
			\multicolumn{1}{c}{Channel} &
			\multicolumn{1}{c}{Width} &
			\multicolumn{1}{c}{Branching fraction} \\
			\hline
			\multirow{9}{*}{$1F(2^{+},5/2)$} 
			& $D^{*+}\pi^-$    &366.04 $g_{ZH}^2$  & 13.11
			& $D_{s}^{*+}K^0$      &489.56 $g_{ZH}^2$  &10.68 \\ 
			& $D^{*0}\pi^0$    &186.79 $g_{ZH}^2$  & 6.69
			& $D_{s}^{*0}K^+$      &502.58 $g_{ZH}^2$  & 10.96\\ 
			& $D^{*0}\eta$     &29.30 $g_{ZH}^2$  & 1.05
			& $D_{s}^{*+}\eta$     &161.73 $g_{ZH}^2$  &3.53 \\ 
			& $D_{s}^{*+}K^-$  &101.27 $g_{ZH}^2$  &3.63
			& $D_{s}^{*+}\pi^0$    &0.02 $g_{ZH}^2$  &0.00 \\ 
			& $D^{+}\pi^-$    &1060.71 $g_{ZH}^2$  & 37.99
			& $D_{s}^{+}K^0$      &1423.80 $g_{ZH}^2$  &31.06 \\ 
			& $D^{0}\pi^0$    &541.978$g_{ZH}^2$  & 19.41
			& $D_{s}^{0}K^+$      &1460.77 $g_{ZH}^2$  & 31.87\\ 
			& $D^{0}\eta$     &104.96 $g_{ZH}^2$  & 3.76
			& $D_{s}^{+}\eta$     &545.06 $g_{ZH}^2$  &11.89 \\ 
			& $D_{s}^{+}K^-$  &400.72 $g_{ZH}^2$  &14.35
			& $D_{s}^{+}\pi^0$    &0.068 $g_{ZH}^2$  &0.00 \\ 
			& \textbf{Total}   &2791.79   $g_{ZH}^2$  &
			& \textbf{Total}   &4583.61   $g_{ZH}^2$  & \\
			\hline
			\multirow{5}{*}{$1F(3^{+},5/2)$} 
			& $D^{*+}\pi^-$    &997.69 $g_{ZH}^2$  & 53.17
			& $D_{s}^{*+}K^0$      &1344.37 $g_{ZH}^2$  & 42.32\\ 
			& $D^{*0}\pi^0$    &508.91 $g_{ZH}^2$  & 27.12
			& $D_{s}^{*0}K^+$      &1379.13 $g_{ZH}^2$  &43.41 \\ 
			& $D^{*0}\eta$     & 82.06 $g_{ZH}^2$ & 4.37
			& $D_{s}^{*+}\eta$     &453.31 $g_{ZH}^2$  &14.27 \\ 
			& $D_{s}^{*+}K^-$  &287.68 $g_{ZH}^2$  &15.33
			& $D_{s}^{*+}\pi^0$    &0.06 $g_{ZH}^2$  &0.00 \\  
			& \textbf{Total}   & 1876.35  $g_{ZH}^2$   &
			& \textbf{Total}   &3176.87  $g_{ZH}^2$    & \\
			\hline
			\multirow{5}{*}{$1F(3^{+},7/2)$} 
			& $D^{*+}\pi^-$    &1921.25 $g_{RH}^2$  & 60.87
			& $D_{s}^{*+}K^0$      &2228.16 $g_{RH}^2$  & 44.18\\ 
			& $D^{*0}\pi^0$    &990.30 $g_{RH}^2$  & 31.38
			& $D_{s}^{*0}K^+$      &2327.56 $g_{RH}^2$  & 46.15\\ 
			& $D^{*0}\eta$     &63.34 $g_{RH}^2$  & 2.01
			& $D_{s}^{*+}\eta$     &487.34 $g_{RH}^2$  & 9.66\\ 
			& $D_{s}^{*+}K^-$  &181.25 $g_{RH}^2$  &5.74
			& $D_{s}^{*+}\pi^0$    &0.01 $g_{RH}^2$  &0.00 \\  
			& \textbf{Total}   &3156.15 $g_{RH}^2$  &
			& \textbf{Total}   &5043.08 $g_{RH}^2$  & \\
			\hline
			\multirow{9}{*}{$1F(4^{+},7/2)$} 
			& $D^{*+}\pi^-$    &1154.12 $g_{RH}^2$  & 19.79
			& $D_{s}^{*+}K^0$      &1358.00 $g_{RH}^2$  & 14.22\\ 
			& $D^{*0}\pi^0$    &594.61 $g_{RH}^2$  & 10.20
			& $D_{s}^{*0}K^+$      &1417.35 $g_{RH}^2$  &14.84 \\ 
			& $D^{*0}\eta$     & 39.56 $g_{RH}^2$ & 0.68
			& $D_{s}^{*+}\eta$     &304.21 $g_{RH}^2$  & 3.19\\ 
			& $D_{s}^{*+}K^-$  &114.89 $g_{RH}^2$  &1.97
			& $D_{s}^{*+}\pi^0$    &0.09 $g_{RH}^2$  & 0.00\\ 
			& $D^{+}\pi^-$    &2253.29 $g_{RH}^2$  & 38.64
			& $D_{s}^{+}K^0$      & 2782.71 $g_{RH}^2$ & 29.14\\ 
			& $D^{0}\pi^0$    &1162.47 $g_{RH}^2$  & 19.94
			& $D_{s}^{0}K^+$      &2898.58 $g_{RH}^2$  & 30.36\\ 
			& $D^{0}\eta$     &117.86 $g_{RH}^2$  & 2.02
			& $D_{s}^{+}\eta$     &787.22 $g_{RH}^2$  & 8.24\\ 
			& $D_{s}^{+}K^-$  &394.36 $g_{RH}^2$  &6.76
			& $D_{s}^{+}\pi^0$    &0.17 $g_{RH}^2$  & 0.00\\ 
			& \textbf{Total}   & 5831.18 $g_{RH}^2$ &
			& \textbf{Total}   &9548.34 $g_{RH}^2$  & \\
		\end{tabular}
	\end{ruledtabular}
\end{table*}
\end{widetext}

\section{STRONG DECAY}
%

{Beyond the mass spectrum, the strong-decay pattern of the $D$ and $D_s$ families provides an essential test of state assignments. To analyze these decays, we employ heavy quark effective theory (HQET) framework combined with chiral symmetry. In this framework, charmed mesons are arranged into heavy-quark spin doublets, and their couplings to the light pseudoscalar mesons ($\pi$, $K$, $\eta$) are described by a leading-order chiral effective Lagrangian that preserves heavy-quark spin–flavor symmetry. Further details of the formalism and its implementation can be found in \cite{Falk:1992cx, Falk:1995th}. From this effective Lagrangian, one can derive explicit formulae for the two-body strong decay widths. These expressions depend on a set of dimensionless strong coupling constants (denoted $g_{HH}$, $g_{SH}$, $g_{TH}$, etc.), which parametrize the strength of transitions between specific heavy-meson multiplets and the light meson field.}
{At leading order, the two-body strong decay widths of  various charmed meson states decaying to another charmed meson states and a light pseudo scalar meson($\mathcal{P}$=$\pi$, $\eta$, $K$) are given by the following expressions \cite{Falk:1992cx, Falk:1995th, Gupta:2018zlg}:}
\begin{enumerate}
\item {For $S$-wave doublet ($J_i^P = 0^-, 1^-$) to $1S$ ground-state ($J_f^P = 0^-, 1^-$) +  $\mathcal{P}$:}
$$
\Gamma(1^- \to 0^-) \;=\; C_\mathcal{P} \, \frac{g_{HH}^2}{6\pi f_\pi^2}\, \frac{M_f}{M_i}\, P_{\mathcal{P}}^3
$$
$$
\Gamma(1^- \to 1^-) \;=\; C_\mathcal{P} \, \frac{g_{HH}^2}{3\pi f_\pi^2}\, \frac{M_f}{M_i}\, P_{\mathcal{P}}^3
$$
$$
\Gamma(0^- \to 1^-) \;=\; C_\mathcal{P} \, \frac{g_{HH}^2}{2\pi f_\pi^2}\, \frac{M_f}{M_i}\, P_{\mathcal{P}}^3
$$

\item {For $j=\tfrac{1}{2}$ $P$-wave doublet ($J_i^P = 0^+, 1^+$) to $1S$ ground-state ($J_f^P = 0^-, 1^-$) + $\mathcal{P}$:}
$$
\Gamma(0^+ \to 0^-) \;=\; C_\mathcal{P} \, \frac{g_{SH}^2}{2\pi f_\pi^2}\, \frac{M_f}{M_i}
\left[ M_{\mathcal{P}}^2 + P_{\mathcal{P}}^2 \right] P_{\mathcal{P}}
$$
$$
\Gamma(1^+ \to 1^-) \;=\; C_\mathcal{P} \, \frac{g_{SH}^2}{2\pi f_\pi^2}\, \frac{M_f}{M_i}
\left[ M_{\mathcal{P}}^2 + P_{\mathcal{P}}^2 \right] P_{\mathcal{P}}
$$

\item {For $j=\tfrac{3}{2}$ $P$-wave doublet ($J_i^P = 1^+, 2^+$) to $1S$ ground-state ($J_f^P = 0^-, 1^-$) + $\mathcal{P}$:}

$$
\Gamma(1^+ \to 1^-) \;=\; C_\mathcal{P} \, \frac{2 g_{TH}^2}{3\pi f_\pi^2 \Lambda^2}\, \frac{M_f}{M_i}\, P_{\mathcal{P}}^5
$$
$$
\Gamma(2^+ \to 0^-) \;=\; C_\mathcal{P} \, \frac{4 g_{TH}^2}{15\pi f_\pi^2 \Lambda^2}\, \frac{M_f}{M_i}\, P_{\mathcal{P}}^5
$$

$$
\Gamma(2^+ \to 1^-) \;=\; C_\mathcal{P} \, \frac{2 g_{TH}^2}{5\pi f_\pi^2 \Lambda^2}\, \frac{M_f}{M_i}\, P_{\mathcal{P}}^5
$$

\item {For $j=\tfrac{3}{2}$ $D$-wave doublet ($J_i^P = 1^-, 2^-$) to $1S$ ground-state ($J_f^P = 0^-, 1^-$) + $\mathcal{P}$}

$$
\Gamma(1^- \to 0^-) \;=\; C_\mathcal{P} \, \frac{4 g_{XH}^2}{9\pi f_\pi^2 \Lambda^2}\, \frac{M_f}{M_i}
\left[ M_{\mathcal{P}}^2 + P_{\mathcal{P}}^2 \right] P_{\mathcal{P}}^3
$$

$$
\Gamma(1^- \to 1^-) \;=\; C_\mathcal{P} \, \frac{2 g_{XH}^2}{9\pi f_\pi^2 \Lambda^2}\, \frac{M_f}{M_i}
\left[ M_{\mathcal{P}}^2 + P_{\mathcal{P}}^2 \right] P_{\mathcal{P}}^3
$$

$$
\Gamma(2^- \to 1^-) \;=\; C_\mathcal{P} \, \frac{2 g_{XH}^2}{3\pi f_\pi^2 \Lambda^2}\, \frac{M_f}{M_i}
\left[ M_{\mathcal{P}}^2 + P_{\mathcal{P}}^2 \right] P_{\mathcal{P}}^3
$$

\item {For $j=\tfrac{5}{2}$ $D$-wave doublet ($J_i^P = 2^-, 3^-$) to $1S$ ground-state ($J_f^P = 0^-, 1^-$) + $\mathcal{P}$:}
$$
\Gamma(2^- \to 1^-) \;=\; C_\mathcal{P} \, \frac{4 g_{YH}^2}{15\pi f_\pi^2 \Lambda^4}\, \frac{M_f}{M_i}\, P_{\mathcal{P}}^7
$$

$$
\Gamma(3^- \to 0^-) \;=\; C_\mathcal{P} \, \frac{4 g_{YH}^2}{35\pi f_\pi^2 \Lambda^4}\, \frac{M_f}{M_i}\, P_{\mathcal{P}}^7
$$

$$
\Gamma(3^- \to 1^-) \;=\; C_\mathcal{P} \, \frac{16 g_{YH}^2}{105\pi f_\pi^2 \Lambda^4}\, \frac{M_f}{M_i}\, P_{\mathcal{P}}^7
$$

\item {For $j=\tfrac{5}{2}$ $F$-wave doublet ($J_i^P = 2^+, 3^+$) to $1S$ ground-state ($J_f^P = 0^-, 1^-$) + $\mathcal{P}$:}
$$
\Gamma(2^+ \to 0^-) \;=\; C_\mathcal{P} \, \frac{4 g_{ZH}^2}{25\pi f_\pi^2 \Lambda^4}\, \frac{M_f}{M_i}
\left[ M_{\mathcal{P}}^2 + P_{\mathcal{P}}^2 \right] P_{\mathcal{P}}^5
$$

$$
\Gamma(2^+ \to 1^-) \;=\; C_\mathcal{P} \, \frac{8 g_{ZH}^2}{75\pi f_\pi^2 \Lambda^4}\, \frac{M_f}{M_i}
\left[ M_{\mathcal{P}}^2 + P_{\mathcal{P}}^2 \right] P_{\mathcal{P}}^5
$$

$$
\Gamma(3^+ \to 1^-) \;=\; C_\mathcal{P} \, \frac{4 g_{ZH}^2}{15\pi f_\pi^2 \Lambda^4}\, \frac{M_f}{M_i}
\left[ M_{\mathcal{P}}^2 + P_{\mathcal{P}}^2 \right] P_{\mathcal{P}}^5
$$

\item {For $j=\tfrac{7}{2}$ $F$-wave doublet ($J_i^P = 3^+, 4^+$) to $1S$ ground-state ($J_f^P = 0^-, 1^-$) + $\mathcal{P}$:}
$$
\Gamma(3^+ \to 1^-) \;=\; C_\mathcal{P} \, \frac{36 g_{RH}^2}{35\pi f_\pi^2 \Lambda^6}\, \frac{M_f}{M_i}\, P_{\mathcal{P}}^9
$$

$$
\Gamma(4^+ \to 0^-) \;=\; C_\mathcal{P} \, \frac{16 g_{RH}^2}{35\pi f_\pi^2 \Lambda^6}\, \frac{M_f}{M_i}\, P_{\mathcal{P}}^9
$$

$$
\Gamma(4^+ \to 1^-) \;=\; C_\mathcal{P} \, \frac{4 g_{RH}^2}{7\pi f_\pi^2 \Lambda^6}\, \frac{M_f}{M_i}\, P_{\mathcal{P}}^9
$$

{In the above formulas, $C_{\mathcal{P}}$ is a coefficient that depends on the identity of the emitted pseudoscalar meson (essentially an isospin factor). Specifically, $C_{\mathcal{P}} = 1$ for $\pi^\pm$, $K^\pm$, or $K^0/\bar K^0$; $C_{\mathcal{P}} = \tfrac{1}{2}$ for $\pi^0$; and for $\eta$ emission $C_{\mathcal{P}}$ takes the value $\tfrac{1}{6}$ if the decaying meson is a non-strange $D$, or $\tfrac{2}{3}$ if it is a $D_s$. In each decay, $M_i$ and $M_f$ denote the masses of the initial and final meson, respectively, while $P_{\mathcal{P}}$ is the momentum of the outgoing pseudoscalar meson (with $M_{\mathcal{P}}$ being the pseudoscalar’s mass). We set the chiral symmetry-breaking scale $\Lambda = 1~\text{GeV}$ and $f_{\pi}=132~\text{MeV}$.}
{We use different symbols for the strong couplings depending on the multiplets involved and the radial quantum number of decaying meson. For decays among the lowest-lying states ($n=1$), the couplings are written as $g_{HH}$, $g_{SH}$, $g_{TH}$, etc. For transitions from first excited states ($n=2$) down to the ground state, we use tilded couplings $\tilde g$ (e.g. $\tilde{g}_{HH}, \tilde{g}_{SH}$, …), and similarly a double tilde (such as $\tilde{\tilde g}_{HH}$) denotes the couplings for decays from second radially excited ($n=3$) states to the ground state. 
}

{Using the above expressions, together with our computed masses for the initial and final $D$ and $D_s$ states and the PDG values for the masses of light pseudo scalar meson  \cite{PDG2024}, we calculate the strong decay widths of the relevant charmed mesons. The resulting partial decacy widths are reported in terms of the corresponding strong coupling constants in Table \ref{tbld1}-\ref{tbld3}. 
}
\end{enumerate}

\section{RESULTS AND DISCUSSION}

\subsection{Charmed and charmed strange mesons}
In this section, we present and discuss the results obtained for the mass spectra of charmed ($D$) and charmed-strange ($D_s$) mesons within our theoretical framework. To represent the state of the system, we use the notation $nL(J^P, j)$, where $n = 1, 2, 3, \ldots$ denotes the radial quantum number of the light component, and $L = S, P, D, \ldots$ specifies its orbital angular momentum. This is followed by the total angular momentum $J$ and parity $P$ of the state, and finally, the light component's total angular momentum $j$.  The calculated masses are summarized in Tables \ref{tbl1} and \ref{tbl2}, respectively, and systematically compared with experimental data from PDG \cite{PDG2024} as well as other theoretical predictions available in the literature \cite{Godfrey:2015dva, Ebert:2009ua, Song:2015fha, Song:2015nia}.

\subsubsection{$D$ Meson}

Significant experimental progress has enabled a comprehensive understanding of the $D$-meson spectrum. The states $D^{0,\pm}$ and $D^{*}(2007)^{0}/D^{*}(2010)^{\pm}$, firmly established as the ground-state doublet $1S(0^{-},1/2)$ and $1S(1^{-},1/2)$ respectively, are well documented experimentally \cite{PDG2024}. Our theoretical calculations accurately reproduce their experimentally observed masses.

Regarding the $1P$ states, the PDG \cite{PDG2024} identifies $D_0^{*}(2300)$, $D_1(2420)$, $D_1(2430)^{0}$, and $D_2^{*}(2460)$ as the primary candidates. The neutral $D_0^*(2300)$, with quantum numbers $J^P = 0^+$, was initially observed by Belle  and subsequently confirmed by \textit{BABAR}. The charged counterpart, $D_0^*(2300)^+$, observed by FOCUS, was later confirmed by LHCb. Our computed mass $2362\pm7\,\text{MeV}$ for the $1P(0^{+},1/2)$ state aligns well with the experimental average $2343\pm10\,\text{MeV}$ for $D_0^*(2300)$, consistent also with previous theoretical studies \cite{Song:2015fha, Lu:2014zua}. The predicted masses of other $1P$ states ($2422\pm6\,\text{MeV}$, $2429\pm5\,\text{MeV}$, and $2462\pm4\,\text{MeV}$) show excellent agreement with the measured values for $D_1(2420)$, $D_1(2430)^0$, and $D_2^*(2460)$.

{Taking the decay width of the $D_2^{*}(2460)$, $ 47.3 \pm 0.8$ MeV \cite{PDG2024}, the corresponding coupling constant is fixed to $g_{TH} = 0.413 \pm 0.0035$. With this, the resulting partial-width ratio for $1P(2^{+},3/2)$ is found to be
$
\frac{\Gamma\!\left(D_{1P(2^{+},3/2)}\to D^{+}\pi^{-}\right)}
{\Gamma\!\left(D_{1P(2^{+},3/2)}\to D^{*+}\pi^{-}\right)}
= 2.28 \, ,
$
which also lies within the experimentally measured range reported by ZEUS and CLEO \cite{PDG2024}. For the same state, $D_2^{*}(2460)$, the BaBar collaboration  measured the ratio
$
\frac{\Gamma(D^{+}\pi^{-})}
{\Gamma(D^{+}\pi^{-})+\Gamma(D^{*+}\pi^{-})}
= 0.62 \pm 0.03 \pm 0.02 \, 
$
\cite{PDG2024}. Our corresponding estimate yields a value of $0.69$ for this ratio, indicating reasonable consistency with the BaBar result.}

Several higher excitations, including the states $D(2550)^0$, $D^*(2600)^{0,+}$, $D(2750)^0$, and $D^*(2760)^{0,+}$, were first observed by BABAR and subsequently confirmed by LHCb \cite{PDG2024}. A detailed four-body amplitude analysis of the $ B^- \to D^{*+} \pi^- \pi^- $ decay  by LHCb \cite{PDG2024} categorized these $D(2550)$, $ D^*(2600) $, $D(2750)$, and $ D^*(2760) $ states as $J^P = 0^-$, $1^-$, $2^-$, and $3^-$, respectively.
These determinations for $D^*(2600)$ and $D^*(2760)$ are in agreement with earlier experimental analyses \cite{PDG2024}.
The Particle Data Group (PDG) labels these states $D(2550)$, $D^*(2600)$, $D(2750)$, and $D^*(2760)$ as $D_0(2550)^{0}$, $D^*_1(2600)^{0}$, $D_2(2740)^{0}$, and $D^*_3(2750)$, respectively \cite{PDG2024}.

Our calculated mass $2754\pm4$ MeV for \(1D(3^{-},5/2)\) is in excellent agreement with experimental average mass  $2763.1\pm3.2$ MeV for \( D^*_3(2750) \). 
{The experimentally measured width of the $D_3^{*}(2750)$, $\Gamma = 66 \pm 5~\mathrm{MeV}$, is used to extract the corresponding strong coupling, yielding $g_{YH} = 0.404 \pm 0.015$. Further, the partial-width ratio
$
	R_\pi\!\left(D_{1D(3^{-},\,5/2)}\right)
	=
	\frac{\Gamma\!\left(D_{1D(3^{-},\,5/2)}\to D\pi\right)}
	{\Gamma\!\left(D_{1D(3^{-},\,5/2)}\to D^{*}\pi\right)}
	= 1.97
$
is obtained. This result exceeds the value $0.42 \pm 0.05 \pm 0.11$ reported by the BaBar Collaboration in 2010 \cite{PDG2024}. This suggests that additional measurements from other experimental facilities are also needed to better establish the decay pattern of the $D_3^{*}(2750)$.
}

The resonance $D_1^*(2760)^0$, experimentally observed at $2781\pm22\,\text{MeV}$ with $J^{P}=1^{-}$ by LHCb \cite{PDG2024}, closely matches our prediction for the $1D(1^-,3/2)$ state ($2739\pm5\,\text{MeV}$).  Although the predicted and experimental uncertainty ranges do not completely overlap, the non-overlapping region is only about \( 15 \, \text{MeV} \), which remains acceptable within the typical uncertainties associated with heavy meson mass predictions. Therefore, the \( D_1^*(2760)^0 \) resonance may be interpreted as the \( 1D(1^-,3/2) \) state. This assignment is also consistent with theoretical predictions given in Refs.\cite{Wang:2010ydc, Lu:2014zua, Song:2015fha, Chen:2015lpa, Godfrey:2015dva}. 
{Moreover, the dominant decay mode of the $1D(1^{-},\,3/2)$ state is predicted to be $D\pi$(as shown in the table \ref{tbld1} listing the decay channels of the $1D(1^{-},\,3/2)$ state), which aligns with the experimental observation of the $D_1^{*}(2760)^0$ in this channel. The LHCb Collaboration reports a total decay width of $177 \pm 32 \pm 21\,\mathrm{MeV}$ for this state \cite{PDG2024}. Using this measured width as input, we extract the corresponding strong coupling and obtain $g_{XH} = 0.253 \pm 0.027$, further supporting the proposed spectroscopic assignment.}

{The measured mass of the $D_2(2740)^0$ state, $2747 \pm 6$ MeV, lies within the uncertainty range of our predicted mass for the $1D(2^{-},\,5/2)$ configuration. In contrast, the experimental mass is located approximately $6$ MeV outside the uncertainty band of the calculated mass for the $1D(2^{-},\,3/2)$ state. Using the previously extracted values of the couplings $g_{XH}$ and $g_{YH}$, we obtain decay widths of $92 \pm 23\,\mathrm{MeV}$ and $32.5 \pm 2.5\,\mathrm{MeV}$ for the $1D(2^{-},\,3/2)$ and $1D(2^{-},\,5/2)$ states, respectively. Given the experimentally measured width of the $D_2(2740)^0$, $\Gamma = 88 \pm 19\,\mathrm{MeV}$ \cite{PDG2024}, the assignment of this state as $1D(2^{-},\,3/2)$ is therefore favored.}

Taken together, these results support the $J^{P}$ assignments proposed by the LHCb Collaboration, identifying the $D_3^{*}(2750)$ and $D_2(2740)^0$ states as having quantum numbers $3^{-}$ and $2^{-}$, respectively \cite{PDG2024}. These assignments are also consistent with the theoretical predictions reported in Refs. \cite{Wang:2010ydc, Lu:2014zua, Song:2015fha, Godfrey:2015dva}.

{In the case of radially excited states, the calculated masses for the $2S(0^-,1/2)$ and $2S(1^-,1/2)$ states ($2594\pm6\,\text{MeV}$ and $2597\pm6\,\text{MeV}$) are proximate to experimentally observed $D_0(2550)^{0}$ and $D^*_1(2600)^{0}$ resonances. Despite minor deviations of about $30-45\,\text{MeV}$, $D_0(2550)^{0}$ and $D^*_1(2600)^{0}$ states remain excellent candidates for radial excitations, consistent with previous theoretical analyses \cite{Li:2010vx, Wang:2010ydc, Yu:2014dda, Lu:2014zua}.
Employing our extracted value of the coupling $\tilde g_{HH} = 0.264 \pm 0.011$ (the details of its determination are discussed in the subsequent section for the $D_s$ sector), we compute the decay width of the $D(2S(1^{-},\,1/2))$ state to be $141 \pm 23\,\mathrm{MeV}$. This result compares favorably with the experimentally measured width of $165 \pm 24\,\mathrm{MeV}$ \cite{PDG2024} for the $D_1^{*}(2600)^0$, lending further support to its interpretation as the $2S(1^{-},\,1/2)$ excitation. Using the same coupling, the decay width of the $D(2S(0^{-},\,1/2))$ state is predicted to be $107 \pm 17\,\mathrm{MeV}$. While this value is somewhat smaller than the experimental width quoted for the $D_0(2550)^0$, $\Gamma = 165 \pm 24\,\mathrm{MeV}$ \cite{PDG2024}, the difference remains moderate in view of the sizable experimental uncertainty.  Further insight is provided by the decay pattern: our calculation yields a dominant branching fraction of approximately $65.9\%$ into the $D^{*}\pi^{-}$ channel for the $2S(0^{-},\,1/2)$ state. This behavior is consistent with experimental observations, as the $D_0(2550)^0$ has been reported in the $D^{*}\pi^{-}$ final state by the LHCb Collaborations  and by BaBar \cite{PDG2024}. Taken together, the mass spectrum, decay widths, and branching structure support the assignment of the $D_0(2550)^0$ as the $2S(0^{-},\,1/2)$ charmed meson.}

{The LHCb Collaboration has reported the observation of two higher excited charmed meson states. One of these is a natural-parity state, denoted $D^{*}_{J}(3000)$, observed with a mass of $3008.1 \pm 4.0\,\mathrm{MeV}$ and a total decay width of $110.5 \pm 11.5\,\mathrm{MeV}$, predominantly decaying into the $D\pi$ final state. In addition, an unnatural-parity resonance, labeled $D_{J}(3000)$, has been identified with a measured mass of $2971.8 \pm 8.7\,\mathrm{MeV}$. This state is observed mainly in the $D^{*}\pi$ channel and exhibits a comparatively larger width of $188.1 \pm 44.8\,\mathrm{MeV}$ \cite{PDG2024}.
Our theoretical mass predictions for the $3S$ and $1F$ multiplets are found to be in close proximity to this mass region.
So, the predicted masses suggests that the $D^*_J(3000)$ state is most plausibly associated with one of the following natural parity states: $3S\,(1^{-},\,1/2)$, $1F\,(2^{+},\,5/2)$, or $1F\,(4^{+},\,7/2)$. For the $D_J(3000)$ state, the most probable assignments from the predicted masses include the $3S\,(0^{-},\,1/2)$,  $1F\,(3^{+},\,/2)$  and $1F\,(3^{+},\,7/2)$ configurations. 
As the $D_J^*(3000)$ state has been reported in the $D\pi$ channel,  the ratio
$R_\pi\!\left(D_{3S\,(1^{-},\,1/2)}\right)=\dfrac{\Gamma\!\left(D_{3S\,(1^{-},\,1/2)}\to D\pi\right)}{\Gamma\!\left(D_{3S\,(1^{-},\,1/2)}\to D^*\pi\right)}=0.64$
disfavour  $3S\,(1^{-},\,1/2)$ assignment for  $D_J^*(3000)$ state. Further if $D_J(3000)$ is assigned as the $3S\,(0^{-},\,1/2)$ state, we extract the coupling
$\tilde{\tilde g}_{HH}=0.163 \pm 0.019$.
The $1F(2^{+},\,5/2)$ and $1F(4^{+},\,7/2)$ states are also possible choices for $D_J^{*}(3000)$. Both mainly decay to $D\pi$. We find
$R_\pi\!\left(D_{1F(2^{+},\,5/2)}\right)=2.90,$ and $R_\pi\!\left(D_{1F(4^{+},\,7/2)}\right)=1.95.$
Because the $1F(2^{+},\,5/2)$ state gives a larger preference for $D\pi$, we prefer $D_J^{*}(3000)$ to be $1F(2^{+},\,5/2)$. Using the experimental width of $D_J^{*}(3000)$ state, we extract
$
g_{ZH}=0.199\pm0.010.
$
Overall, our results support $D_J^*(3000)$ as most likely the $1F(2^{+},5/2)$ state, while the $1F(4^{+},7/2)$ assignment for $D_J^*(3000)^0$ cannot be ruled out.
For $D_J(3000)$, the $1F(3^{+},\,5/2)$ and $1F(3^{+},\,7/2)$ assignments are still reasonable because they mostly decay to $D^*\pi$, with
$
BR(D^*\pi)\approx 80\text{--}90\%.
$
The decay to $D_sK$ can help to test these assignments. We predict
$
BR(D_sK)\simeq 14\text{--}15\% \quad \text{for } 1F(2^{+},\,5/2)\ \text{and}\ 1F(3^{+},\,5/2),
$
and
$
BR(D_sK)\simeq 6\text{--}7\% \quad \text{for } 1F(3^{+},\,7/2)\ \text{and}\ 1F(4^{+},\,7/2).
$
Future measurements of these modes can clarify the nature of both $D_J(3000)$ and $D_J^{*}(3000)$.}

Lastly, the $D^*_2(3000)$ state was observed by the LHCb Collaboration with a measured mass of $3214 \pm 29 \pm 49 \,\text{MeV}$, decay width $186 \pm 38~\mathrm{MeV}$, and quantum numbers $J^P = 2^+$ \cite{PDG2024}. Our theoretical mass predictions for the $3P$ and $2D$ multiplets place lies in the same mass region. However, given that the $D^*_2(3000)$ state has $J^P = 2^+$, it cannot originate from the $2D$ multiplet. In contrast, our calculated mass for the $3P\,(2^{+},\,3/2)$ state is in excellent agreement with the experimentally measured mass of the $D^*_2(3000)$. Based on this correspondence, we conclude that the $D^*_2(3000)$ state is most likely the $3P\,(2^{+},\,3/2)$ state.

\subsubsection{$D_{s}$ Meson}

Turning to the $D_s$-meson spectrum, the well-established ground-state resonances $D_s^\pm$ and $D_s^{*\pm}$ correspond well with our theoretical masses for the $1S(0^-,1/2)$ and $1S(1^-,1/2)$ states, respectively.

The classification of $D_{s1}(2536)^{\pm}$ and $D_{s2}(2573)^{\pm}$ resonances as the $1P(1^+,3/2)$ and $1P(2^+,3/2)$ states is widely accepted, and our model reproduces their masses accurately. {Using the coupling $g_{TH} = 0.413 \pm 0.0035$ fixed from the $D$ meson sector, we obtain a total decay width of $19.9\,\mathrm{MeV}$ for $1P(2^+,3/2)$ state, which is close to the PDG average value of $16.9 \pm 0.7\,\mathrm{MeV}$. Our predicted partial-width ratio,
$
\frac{\Gamma\!\left(D_{s2}^{*}(2573)\to D^{*0}K^{+}\right)}
{\Gamma\!\left(D_{s2}^{*}(2573)\to D^{0}K^{+}\right)}
= 0.09 \, ,
$
is also consistent with the experimental constraint
$
\frac{\Gamma\!\left(D_{s2}^{*}(2573)\to D^{*0}K^{+}\right)}
{\Gamma\!\left(D_{s2}^{*}(2573)\to D^{0}K^{+}\right)}
< 0.33 \, .
$}

However, significant discrepancies persist for the other two $1P$ states, $D_{s0}(2317)^{\pm}$ and $D_{s1}(2460)^{\pm}$. Our calculations yield predicted masses of $2461 \pm 7$ MeV and $2524 \pm 6$ MeV for these states, respectively. These theoretical values exceed the experimentally measured masses of the $D_{s0}(2317)^{\pm}$ and $D_{s1}(2460)^{\pm}$ by approximately 144 MeV and 64 MeV, respectively. Similar deviations have been consistently observed in various theoretical works \cite{Ebert:2009ua, Godfrey:2015dva, Song:2015nia}.

The positive-parity resonances $D_{s0}(2317)^{+}$ and $D_{s1}(2460)^{+}$, first observed by the \textit{BABAR} and \textit{CLEO}  collaborations, were initially identified as the $1P(0^{+})$ and $1P(1^{+})$ states of $D_{s}$ meson, respectively. However, the significantly lower experimental masses of these states, relative to predictions from conventional quark models \cite{Godfrey:1985xj, DiPierro:2001dwf, Godfrey:2015dva}, present notable theoretical challenges. Comparable discrepancies are also observed in lattice QCD calculations, which consistently predict substantially higher masses than those measured experimentally \cite{Moir:2013ub}. To address this discrepancy—often referred to as the “low-mass puzzle”—various non-conventional interpretations have been proposed. These include the possibility that $D_{s0}(2317)^{\pm}$ and $D_{s1}(2460)^{\pm}$ are exotic configurations such as $D^{(*)}K$ molecular states \cite{Barnes:2003dj, Chen:2004dy, Guo:2006fu, Rosner:2006vc, Faessler:2007us, Ortega:2016mms} or compact tetraquark states \cite{Cheng:2003kg, Browder:2003fk, Dmitrasinovic:2005gc, Bracco:2005kt, Hayashigaki:2004st}. More recent theoretical investigations suggest that                                                                                                                                                                                                                                                                                                                  these resonances may not be pure $1P$ $c\bar{s}$ mesons, but rather mixed states composed of a bare $1P$ $c\bar{s}$ core coupled dynamically to a nearby $D^{(*)}K$ threshold \cite{PhysRevD.94.074037, PhysRevLett.128.112001}. These interpretations point toward a more intricate internal structure for the $D_{s0}(2317)$ and $D_{s1}(2460)$, highlighting the necessity for further experimental measurements and refined theoretical analyses to clarify their nature.

The $D_{sJ}(2860)^+$ resonance was first reported by the \textit{BABAR} collaboration with a measured mass of $M_{\text{exp}} = 2856.6 \pm 6.5$ MeV \cite{PDG2024}. Its existence was subsequently confirmed by the LHCb Collaboration using proton-proton collision data. In 2014, a detailed amplitude analysis by LHCb revealed that the structure observed near this mass consists of two overlapping resonances: $D^*_{s1}(2860)^-$, with quantum numbers $J^P = 1^-$, and $D^*_{s3}(2860)^-$, with $J^P = 3^-$. The $D^*_{s3}(2860)^+$ resonance was later observed in the $D^{*+} K_s^0$ decay channel, with resonance parameters consistent with $D^*_{s3}(2860)^-$ \cite{PDG2024}. 

Our theoretical mass prediction for the $1D\,(1^-,\,3/2)$ state is $2857 \pm 5$ MeV, which is in excellent agreement with the experimentally measured mass of the $D^*_{s1}(2860)^{\pm}$ resonance. {Using previously extracted coupling, $g_{XH}=0.253 \pm 0.027$, we obtain a total width of $263 \pm 57~\mathrm{MeV}$ for $1D\,(1^{-},\,3/2)$  state. Within uncertainties, this is compatible with the experimental width $\Gamma = 160 \pm 80~\mathrm{MeV}$ quoted for the $D^{*}_{s1}(2860)^{\pm}$. Taken together, the mass proximity and the width systematics favor  $D^{*}_{s1}(2860)^{\pm}$ state as the $1D\,(1^{-},\,3/2)$ state. Further, for the $D^{*}_{s3}(2860)^{\pm}$, observed mass is consistent with our prediction for the $1D\,(3^{-},\,5/2)$ configuration. Employing $g_{YH}=0.404 \pm 0.015$, we find $\Gamma\!\left[1D\,(3^{-},\,5/2)\right] = 77 \pm 6~\mathrm{MeV}$, close to the experimental value $53 \pm 10~\mathrm{MeV}$ for the $D^{*}_{s3}(2860)^{\pm}$. The combined agreement in mass and width supports the assignment of $D^{*}_{s3}(2860)^{\pm}$ to the $1D\,(3^{-},\,5/2)$ state.
These identifications place $D^{*}_{s1}(2860)^{\pm}$ and $D^{*}_{s3}(2860)^{\pm}$ as the $J^{P}=1^{-}$ and $J^{P}=3^{-}$ members of the $1D$ $c\bar{s}$ multiplet, consistent with earlier analyses \cite{Wang:2014jua, Song:2014mha, Godfrey:2015dva}.}

For the first radial excitations, the \( D_{s1}^{*}(2700)^{\pm} \) resonance is generally considered to be the  $D_s(2S(1^{-},1/2))$ state. This resonance was first observed by BABAR, later it was confirmed by Belle, and its spin-parity numbers were determined to be \( J^P = 1^- \) \cite{PDG2024}. Later the existence of \( D^*_{s1}(2700)^+ \) was further confirmed using \( pp \) collision data at LHCb. We observe that the measured mass of \( D_{s1}^{*}(2700)^{\pm} \) resonance is well reproduced in our work.
{For the $D_s\!\left(2S(1^{-},\,1/2)\right)$ state we obtain
$
\frac{\Gamma\!\left(D_{s(2S(1^{-},1/2))}\to D^{*}K\right)}{\Gamma\!\left(D_{s(2S(1^{-},1/2))}\to DK\right)}
= 0.92 \, ,
$  
which matches well the BaBar measurement $
\frac{\Gamma\!\left(D_{s1}^*(2700)\to D^{*}K\right)}
{\Gamma\!\left(D_{s1}^*(2700)\to DK\right)}
=0.91 \pm 0.13 \pm 0.12 \, ,
$ \cite{PDG2024}.
Motivated by this agreement, we use the measured total width of $D_{s1}^{*}(2700)^{\pm}$ as input to fix the corresponding coupling $\tilde g_{HH}=0.264 \pm 0.011$.}

The LHCb Collaboration recently reported the observation of a new excited $D_s$ meson, $D_{s0}(2590)^+$, with quantum numbers $J^P = 0^-$. This state has been proposed as a candidate for the first radial excitation, corresponding to the $2S$ level. In our model, the $2S\,(0^-,\,1/2)$ state is predicted to lie at $2709 \pm 7$ MeV, which is consistent with the mass range reported in other theoretical studies \cite{Ebert:2002ig, Godfrey:2015dva}. However, our predicted mass is approximately 100 MeV higher than the experimentally measured mass of the $D_{s0}(2590)^+$ resonance.{ Using the coupling $\tilde g_{HH}=0.264 \pm 0.011$, we obtain
$\Gamma\!\left[D_s(2S(0^{-},\,1/2))\right]=80 \pm 7~\mathrm{MeV}$,
which is compatible with the experimental width $89 \pm 20~\mathrm{MeV}$ for $D_{s0}(2590)^+$ \cite{PDG2024}. Even so, the large difference in mass raises questions regarding the identification of the $D_{s0}(2590)^+$ as a conventional $2S\,(0^-,\,1/2)$ state.} 
Therefore, it may be too early to call $D_{s0}(2590)^+$ a standard radial excitation, and more experimental studies are needed. 

Finally, the $D_{sJ}(3040)^+$ resonance was first observed by the \textit{BABAR} Collaboration in the $D^{*}K$ decay channel. Subsequently, the LHCb Collaboration reported weak evidence for the $D_{sJ}(3040)^+$, indicating that it is consistent with an unnatural parity assignment  \cite{PDG2024}. The measured mass of the $D_{sJ}(3040)^+$ is $3044^{+31}_{-9}$ MeV, which aligns well with our theoretical predictions for the $2P$ multiplet. In particular, this multiplet includes the unnatural parity states $2P\,(1^+,\,1/2)$ and $2P\,(1^+,\,3/2)$. Given the agreement between the experimental mass and our calculated values, we suggest that the $D_{sJ}(3040)^+$ resonance may correspond to a $2P$ excitation of the $D_s$ meson family with quantum numbers $J^P = 1^+$. This interpretation is also supported by previous theoretical studies  \cite{Chen:2009zt, Song:2015nia, Godfrey:2015dva}. These comprehensive predictions thus provide critical benchmarks for ongoing and future experimental explorations.

\subsection{Doubly charmed baryons}

The calculated masses for the $\Xi_{cc}$ and $\Omega_{cc}$  baryons are shown in Tables \ref{tbl3} and \ref{tbl4}, respectively. 
{The states listed in Tables \ref{tbl3} and \ref{tbl4} are labeled by $n_d L_d\, n L(J^P,j)$. Here $n_d=1,2,3,\ldots$ counts the radial excitation inside the $cc$ diquark, i.e., the excitation of one $c$ quark relative to the other, while $n=1,2,3,\ldots$ specifies the radial excitation of the light quark with respect to the diquark as a whole. The symbol $L_d=S,P,D,\ldots$ denotes the internal orbital angular momentum of the two $c$ quarks within the diquark, whereas $L=s,p,d,\ldots$ refers to the orbital angular momentum associated with the light-quark motion relative to the diquark. The baryon’s total angular momentum and parity are indicated by $J$ and $P$, and $j$ represents the total angular momentum carried by the light degrees of freedom relative to the $cc$ diquark.}

The only experimentally confirmed baryon containing two heavy quarks is the ground state of the doubly charmed $\Xi_{cc}$ baryon. The Particle Data Group (PDG) lists two isospin partners in this sector: $\Xi_{cc}^{+}$ and $\Xi_{cc}^{++}$. The $\Xi_{cc}^{+}$ was initially reported by the SELEX Collaboration in the $\Lambda_c^{+}K^{-}\pi^{+}$ decay channel, with a measured mass of approximately 3519 MeV \cite{PDG2024}. However, subsequent searches by BaBar, Belle, and LHCb failed to confirm the existence of this state, leading the PDG to classify it with a one-star rating, indicating weak experimental evidence. In contrast, the $\Xi_{cc}^{++}$ was unambiguously observed by the LHCb Collaboration through multiple decay channels, with a precisely measured mass of 3621 MeV \cite{PDG2024}. This state is assigned a three-star rating in the PDG and is most naturally interpreted as the ground-state $\Xi_{cc}$ baryon, with quantum numbers $J^P = \tfrac{1}{2}^+$. Our theoretical calculations well reproduce its observed mass. To date, no $\Omega_{cc}$ has been observed experimentally. However, they have been studied using various theoretical methods. We list these predictions in Table  \ref{tbl3} and \ref{tbl4}  for comparison.

Ebert et al. \cite{Ebert:2002ig} analyzed the masses of the $1S$ and $1P$ states of doubly heavy baryons utilizing a relativistic quark model. In contrast, Oudichhya et al. \cite{Oudichhya:2022ssc} derived mass spectra for doubly heavy baryons employing Regge phenomenology, specifically targeting higher orbital and radial excitations. Non-relativistic quark models were adopted in the studies by Roberts et al. \cite{Roberts:2007ni}, Yoshida et al. \cite{Yoshida:2015tia}, and Valcarce et al. \cite{Valcarce:2008dr} to determine mass spectra for both single- and doubly heavy baryons. Additionally, Giannuzzi \cite{Giannuzzi:2009gh} investigated the $1S$ to $3S$ states within a semi-relativistic quark model framework, leveraging potentials inspired by the AdS/QCD correspondence.

Our computed masses for the $1P$ states of the $\Xi_{cc}$ and $\Omega_{cc}$ baryons fall comfortably within the range of values reported by these varied theoretical approaches, demonstrating a notable agreement with predictions from Refs. \cite{Ebert:2002ig,Oudichhya:2022ssc,Roberts:2007ni,Yoshida:2015tia}. However, we note that our results exceed those presented by Valcarce et al. \cite{Valcarce:2008dr}. For the $1D$ state masses, our calculations align closely with those obtained by Oudichhya et al. \cite{Oudichhya:2022ssc}, while the non-relativistic quark model predictions \cite{Roberts:2007ni,Yoshida:2015tia} appear slightly underestimated. Regarding the $2S$ state masses, our values match well with the predictions by Ref. \cite{Yoshida:2015tia, Giannuzzi:2009gh} but are found to be a bit larger than that predicted by Refs. \cite{Oudichhya:2022ssc, Roberts:2007ni, Valcarce:2008dr}. Further, our predictions for higher orbital excitations ($1F$ and $1G$ states) exhibit good agreement with the Regge phenomenological calculations reported by Oudichhya et al. \cite{Oudichhya:2022ssc}. Overall, our results are mostly consistent with existing theoretical models, although some differences arise due to the specific assumptions used in each approach.

{Moreover, a comparison with the lattice QCD results of Bahtiyar \textit{et al.}~\cite{Bahtiyar:2020uuj} shows that our predictions for the lowest $\Xi_{cc}$ and $\Omega_{cc}$ levels are broadly consistent with first-principles calculations. In particular, Ref.~\cite{Bahtiyar:2020uuj} quotes $1S$ ground-state masses ($J^{P}=\tfrac{1}{2}^{+}$) of $M(\Xi_{cc})=3615 \pm 33\,\mathrm{MeV}$ and $M(\Omega_{cc})=3733 \pm 13\,\mathrm{MeV}$, both of which lie very close to our central values. For the negative-parity sector, their lowest $1P$ excitations with $J^{P}=\tfrac{1}{2}^{-}$ appear at $3930 \pm 20\,\mathrm{MeV}$ for $\Xi_{cc}$ and $4041 \pm 15\,\mathrm{MeV}$ for $\Omega_{cc}$, again compatible with our spectrum. Ref.~\cite{Bahtiyar:2020uuj} further reports a second $\tfrac{1}{2}^{-}$ state nearly degenerate with the first, at about $3971\,\mathrm{MeV}$ for $\Xi_{cc}$ and about $4063\,\mathrm{MeV}$ for $\Omega_{cc}$. When we track these excitations in our framework, we also obtain a set of closely spaced low-lying $\tfrac{1}{2}^{-}$ states, although their absolute positions can differ from the lattice values by roughly $60$--$80\,\mathrm{MeV}$. Overall, the agreement for the ground state and the first excitation is particularly strong, while the remaining differences for higher levels remain moderate, supporting the view that our model reproduces the main spectral features of doubly charmed baryons.}


Furthermore, in our predicted mass spectra for  $\Xi_{cc}$ and $\Omega_{cc}$  baryons, we observe that for states having identical $n$, $L$, and $J$, the mass of the state with higher $j$ is greater than that of the state with lower $j$ for $1S$, $2S$ and $1P$ multiplets. while for higher multiplets we observe that for states having identical $n$, $L$, and $J$, the mass of the state with higher $j$ is lower than that of the state with higher $j$. For $1P$-wave, this feature is also reflected in the mass spectra predicted by \cite{Ebert:2002ig}. The distinct characteristics observed in mass spectra from various theoretical models indicate the necessity for additional experimental measurements to effectively differentiate between them.

\section{CONCLUSION}

We have carried out a comprehensive study of the mass spectra of singly charmed mesons ($D$ and $D_s$) and doubly charmed baryons ($\Xi_{cc}$ and $\Omega_{cc}$) within the relativistic flux-tube model framework. In this model, a charm quark and a light antiquark (for $D$ and $D_s$ mesons) or a tightly bound $\{cc\}$ diquark and a light quark (for $\Xi_{cc}$ and $\Omega_{cc}$ baryons) are connected by a rotating color flux tube that encapsulates the QCD string tension. We incorporate spin-dependent interactions in the $j–j$ coupling scheme to account for hyperfine-structure splitting in the spectra. This approach successfully reproduces the mass of the known $D$ and $D_s$ meson states, and it predicts a rich spectrum of excited states. {To complement the spectroscopy, we also study strong decays of the $D$ and $D_s$ mesons within the heavy quark effective theory (HQET) framework. Using the measured total widths of well-established resonances, we extract the relevant strong couplings, and we find that the resulting couplings provide a consistent description of the decay widths of other experimentally observed charmed-meson states.}

For the charmed mesons, the calculated spectrum and strong decay width allows us to classify experientially observed resonances. Notably, the $D_2(2740)^{0}$ is interpreted as a $1D$ excitation with $J^P=2^-$ (with the light-quark total angular momentum $j=5/2$ or $3/2$), and the $D^*_3(2750)$ is identified as the companion $1D(3^{-},5/2)$ state. The lower-mass resonances $D(2550)^{0}$ and $D^*(2600)^{0}$ are assigned to the first radial excitations , $2S(0^-,1/2)$ and $2S(1^-,1/2)$, respectively, while the neutral $D^*_1(2760)^0$ fits well as the remaining $1D$ $(1^-,3/2)$ state. Furthermore, our model suggests that the broad structures labeled $D_J(3000)$ and $D^*_J(3000)$ correspond to either second radial excitations ($3S$) or members of the first $F$-wave family: specifically,$D^*_J(3000)$ is assigned to:  $1F,(2^{+},5/2)$ or $1F,(4^{+},7/2)$, whereas $D_J(3000)$ is consistent with $3S,(1^{-},1/2)$, $1F,(2^{+},5/2)$, or $1F,(4^{+},7/2)$.
In addition, the newly reported $D^*_2(3000)$ is naturally assigned to the $3P$ $(2^+,j=3/2)$ level. In the charmed-strange meson sector, the observed masses of $D_{s0}^*(2317)^{\pm}$ and $D_{s1}(2460)^{\pm}$ cannot be explained by pure $1P$-wave $c\bar{s}$ states, indicating possible exotic or coupled-channel structures. The $D^*_{s1}(2860)^{\pm}$ and $D*_{s3}(2860)^{\pm}$ are definitively identified as second orbital excitations $1D,(1^-,3/2)$ and $1D,(3^-,5/2)$, respectively, while the $D_{sJ}(3040)^+$ is assigned as a candidate for the $2P$ $(1^+)$ excitation (with $j=1/2$ or $3/2$).

Based on a good description of the well-established states of the charmed meson sector, we have extended the same model to explore the spectrum of doubly charmed baryons. Treating the di-charm $(cc)$ pair as an effective heavy axial-vector diquark in the ground state, attached to a light quark by a flux tube, we predict the masses of orbital and radial excited states of $\Xi_{cc}$ and $\Omega_{cc}$ baryons that have yet to be observed. The mass estimates and quantum number assignments presented here can serve as reference points for upcoming heavy-flavor experiments—especially at LHCb, but also at facilities like Belle II and any future charm factories—in their efforts to discover and identify higher excited charmed mesons and double charmed baryon resonances.

\section{ACKNOWLEDGMENT}
Ms. Pooja Jakhad acknowledges the financial assistance by the Council of Scientific \& Industrial Research (CSIR) under the JRF-FELLOWSHIP scheme with file no. 09/1007(13321)/2022-EMR-I. 
\begin{widetext}
	
\appendix
\section*{APPENDIX: SPIN-DEPENDENT OPERATORS IN HEAVY-LIGHT MESONS AND DOUBLY HEAVY BARYONS}

Here, we detail the construction of the spin-dependent mass-splitting operators and their expectation values for the cases of heavy-light mesons and doubly heavy baryons. We consider the S-wave (ground state) and the first two orbitally excited states (P-wave and D-wave) for demonstration. We start with the $L$–$S$ coupling scheme to construct the basis and then transform it to the $|J,\,j\rangle$ coupling basis. Here, $j$ denotes the total angular momentum of the light degrees of freedom, which is particularly relevant in the heavy-quark limit.

The $L$–$S$ basis states can be constructed using standard Clebsch–Gordan coefficients (CGCs) as follows \cite{Jakhad:2023ids}:

\begin{align}
	|^{2S+1}L_J; J_3\rangle  &= \sum_{S_{l_3},\, S_{h_3},\, L_3,\, S_3} 
	C_{S_{l_3} S_{h_3} S_3}^{S_l S_h S}\, C_{S_3 L_3 J_3}^{S L J}\,|S_{l_3}, S_{h_3}, L_3\rangle, 
	\label{eq:couple-LS-J}
\end{align}

where $S_{h_3}$, $S_{l_3}$, $S_3$, $L_3$, and $J_3$ are the projection quantum numbers corresponding to the quantum numbers $S_h$, $S_l$, $S$, $L$, and $J$, respectively. The state $|S_{h_3},\,S_{l_3},\,L_3\rangle$ denotes the product of the uncoupled states
$|S_h\,S_{h_3}\rangle,|S_l\,S_{l_3}\rangle,$ and $|L\,L_3\rangle.$

The spin-orbit interaction operator $\mathbf{L}\!\cdot\!\mathbf{S}_i$ (for $i = h$ or $l$) can be expressed in terms of ladder operators  as described in Ref.\cite{Karliner:2017kfm}:
\begin{equation}
	\mathbf{L}\!\cdot\!\mathbf{S}_i \;=\; \frac{1}{2}\Big(L_+\,S_{i-} + L_-\,S_{i+}\Big) + L_3\,S_{i3}\,
	\label{eq:LS-operator}
\end{equation}  
which is useful to calculate the expectation value of $\mathbf{L}\!\cdot\!\mathbf{S}_i \;$  by evaluating matrix elements between  $|^{2S+1}L_J\rangle$ basis.

Utilizing these expectation values of $\mathbf{L}\!\cdot\!\mathbf{S}_i \;$  in $|^{2S+1}L_J\rangle$ basis states, we can further calculate the expectation value of the tensor interaction operator using the following relation \cite{Jakhad:2023ids, Karliner:2017kfm}:

\begin{equation}
	\langle\hat{B} \rangle\;=\; -\,\frac{3}{(2L-1)(2L+3)}\Big[\,(\mathbf{L}\!\cdot\!\mathbf{S}_h)(\mathbf{L}\!\cdot\!\mathbf{S}_l)+(\mathbf{L}\!\cdot\!\mathbf{S}_l)(\mathbf{L}\!\cdot\!\mathbf{S}_h)\;-\;\frac{2}{3}\,L(L+1)\,\mathbf{S}_h\!\cdot\!\mathbf{S}_l\,\Big]\,.
	\label{eq:B-tensor}
\end{equation}

Additionally, the expectation value of the spin-spin operator  in $L$–$S$  coupling scheme  can be obtained using the identity $ \mathbf{S} = \mathbf{S}_h + \mathbf{S}_l$ and ${\bf S}^2 = S(S+1)$, one finds 
\begin{equation}
	\langle\mathbf{S}_h\!\cdot\!\mathbf{S}_l\rangle \;=\; \frac{1}{2}\Big[\,S(S+1) - S_h(S_h+1) - S_l(S_l+1)\,\Big]\,
	\label{eq:spinspin-expect}
\end{equation}

\subsection*{A. Heavy–Light Mesons}
\paragraph*{1. The S-wave:} There is no orbital excitation for $L=0$ (S-wave). Only the spin-spin contact hyperfine interaction contributes to the first order. In this case, the total spin $S$ of the meson is $S_h + S_l$, which can be 0 (singlet) or 1 (triplet). The spin-orbit and tensor terms vanish for $L=0$. The expectation value of the spin-spin operator  is the same in either coupling scheme ($L$–$S$ or $J$–$j$) with value $\langle\mathbf{S}_h\!\cdot\!\mathbf{S}_l\rangle=-\frac{3}{4}$ for the spin-singlet ($S=0$) state and $\langle\mathbf{S}_h\!\cdot\!\mathbf{S}_l\rangle=\frac{1}{4}$ for the spin-triplet ($S=1$) state. These reproduce the hyperfine splitting between, e.g., the $D$ meson ($J^P=0^-$, $S=0$) and $D^*$ meson ($J^P=1^-$, $S=1$).

\paragraph*{2. The P-wave:} For $L=1$ (P-wave), the heavy quark ($S_h=\tfrac{1}{2}$) and light antiquark ($S_l=\tfrac{1}{2}$) spins can couple to total spin $S=0$ or $S=1$. These combine with $L=1$ to yield total $J$ quantum numbers $J=0,1,2$ for the meson. In the $L$–$S$ coupling basis, the corresponding states can be labeled as $|{}^1P_1\rangle$ (with $S=0$, $J=1$) and $|{}^3P_J\rangle$ (with $S=1$, $J=0,1,2$). 
In particular, there are two distinct $J=1$ basis states, $|{}^1P_1\rangle$ and $|{}^3P_1\rangle$, which will mix under spin-dependent interactions, while $J=0$ ($|{}^3P_0\rangle$) and $J=2$ ($|{}^3P_2\rangle$) are each unique.

The $L-S$ basis  can be constructed in terms of $|S_{h_3},\,S_{l_3},\,L_3\rangle$ using Eq. [\ref{eq:couple-LS-J} ] as  

\begin{equation} 
	|^{3}P_{0};0\rangle = \frac{1}{\sqrt{3}}|-\frac{1}{2},-\frac{1}{2},1\rangle - \frac{1}{\sqrt{6}}	|\frac{1}{2},-\frac{1}{2},0\rangle- \frac{1}{\sqrt{6}}	|-\frac{1}{2},\frac{1}{2},0\rangle + \frac{1}{\sqrt{3}}	|\frac{1}{2},\frac{1}{2},-1\rangle,
\end{equation}

\begin{equation} 
	|^{1}P_{1};1\rangle = \frac{1}{\sqrt{2}}|\frac{1}{2},-\frac{1}{2},1\rangle - \frac{1}{\sqrt{2}}	|-\frac{1}{2},\frac{1}{2},1\rangle,
\end{equation}

\begin{equation} 
	|^{3}P_{1};1\rangle = -\frac{1}{2}|\frac{1}{2},-\frac{1}{2},1\rangle - \frac{1}{2}	|-\frac{1}{2},\frac{1}{2},1\rangle+ \frac{1}{\sqrt{2}}	|\frac{1}{2},\frac{1}{2},0\rangle,
	\end{equation}
	
	\begin{equation} 
|^{3}P_{2};2\rangle = |\frac{1}{2},\frac{1}{2},1\rangle.
\end{equation}

Using these $L$–$S$ coupling basis, we compute the expectation values of the operators given in Eqs.~[\ref{eq:LS-operator}], [\ref{eq:B-tensor}], and [\ref{eq:spinspin-expect}] within the [$^{2}P_{J}$, $^{4}P_{J}$] basis for all allowed values of $J$, as listed below:\\

For J=0,
\begin{equation}
\text{$\langle\mathbf{L}.\mathbf{S}_l\rangle$=}-1,\ \  
\text{$\langle\mathbf{L}.\mathbf{S}_h\rangle$=}-1,\ \   
\text{$\langle\mathbf{\hat{B}}\rangle$=}-1,\ \ 
\text{$\langle\mathbf{S}_l.\mathbf{S}_h\rangle$=}\frac{1}{4}.
\end{equation}	
For J=1,
\begin{equation}
\text{$\langle\mathbf{L}.\mathbf{S}_l\rangle$=}\left[
\begin{array}{cc}
	0 & -\frac{1}{\sqrt{2}} \\
	-\frac{1}{\sqrt{2}} & -\frac{1}{2} \\
\end{array}
\right],\ \   	
\text{$\langle\mathbf{L}.\mathbf{S}_h\rangle$=}\left[
\begin{array}{cc}
	0 & \frac{1}{\sqrt{2}} \\
	\frac{1}{\sqrt{2}} & -\frac{1}{2} \\
\end{array}
\right],\ \   
\text{$\langle\mathbf{\hat{B}}\rangle$=}\left[
\begin{array}{cc}
	0 & 0 \\
	0 & \frac{1}{2} \\
\end{array}
\right],\ \ 
\text{$\langle\mathbf{S}_l.\mathbf{S}_h\rangle$=}\left[
\begin{array}{cc}
	-\frac{3}{4} & 0 \\
	0 & \frac{1}{4} \\
\end{array}
\right].
\end{equation}	
For J=2,
\begin{equation}
\text{$\langle\mathbf{L}.\mathbf{S}_l\rangle$=}\frac{1}{2},\ \  
\text{$\langle\mathbf{L}.\mathbf{S}_h\rangle$=}\frac{1}{2},\ \   
\text{$\langle\mathbf{\hat{B}}\rangle$=}-\frac{1}{10},\ \ 
\text{$\langle\mathbf{S}_l.\mathbf{S}_h\rangle$=}\frac{1}{4}.
\end{equation}

In the heavy-quark limit ($m_h \gg m_l$), the spin-orbit interaction term involving $\mathbf{L}\cdot\mathbf{S}_l$ dominates over the other interaction terms. Consequently, it is necessary to adopt a basis in which $\mathbf{L}\cdot\mathbf{S}_l$ is diagonal. This condition is satisfied in the $|J, j\rangle$ basis, corresponding to the $j$–$j$ coupling scheme, where the orbital angular momentum $\mathbf{L}$ is first coupled with the light-quark spin $\mathbf{S}_l$ to form $j = L + S_l$. For $L = 1$ and $S_l = \tfrac{1}{2}$, the possible values of $j$ are $\tfrac{1}{2}$ and $\tfrac{3}{2}$. In this limit, meson states can thus be classified by bases $|J, j\rangle$ in which the $\mathbf{L}\cdot \mathbf{S}_l$ matrix is diagonal, while the remaining interaction terms are treated as perturbations.

Diagonalizing the $\mathbf{L}\cdot\mathbf{S}_l$ operator within the $J=1$ subspace, for example, yields two orthogonal linear combinations of $|{}^1P_1>$ and $|{}^3P_1>$ corresponding to definite $j$. The eigenvalues are $\lambda = -1$ for $j=\tfrac{1}{2}$ and $\lambda = +\tfrac{1}{2}$ for $j=\tfrac{3}{2}$ (these values come from $\frac{1}{2}[j(j+1) - L(L+1) - S_l(S_l+1)]$ with $L=1$, $S_l=1/2$). The corresponding eigenstates can be written as:

\begin{equation}
\lambda= -1 :
|J=1, j=\frac{1}{2} \rangle =
\frac{1}{\sqrt{3}}  |^{1}P_{1}   \rangle 
+\sqrt{\frac{2}{3}} |^{3}P_{1}   \rangle,
\end{equation}

\begin{equation}
\lambda= \frac{1}{2} :
|J=1, j=\frac{3}{2} \rangle =
-\sqrt{\frac{2}{3}}  |^{1}P_{1}   \rangle 
+\frac{1}{\sqrt{3}} |^{3}P_{1}   \rangle,
\end{equation}

Similarly, for $J=0$ we have only the $j=\tfrac{1}{2}$ state $|^3P_0\rangle$, and for $J=2$ we have only the $j=\tfrac{3}{2}$ state $|^3P_2\rangle$. These are already eigenstates of $\mathbf{L}\!\cdot\!\mathbf{S}_l$ with eigenvalue $+\tfrac{1}{2}$:
\begin{equation}
|J = 0,\ j = \tfrac{1}{2} \rangle = |^3P_0\rangle,
\end{equation}
\begin{equation}
|J = 2,\ j = \tfrac{3}{2} \rangle = |^3P_2\rangle.
\end{equation}

Having established the $|J,\,j\rangle$ basis (in which $\mathbf{L}\!\cdot\!\mathbf{S}_l$ is diagonal), we can now evaluate the expectation values of the various mass-splitting operators. Table~\ref{table1} summarizes the expectation values of $\mathbf{L}\!\cdot\!\mathbf{S}_h$, $\mathbf{L}\!\cdot\!\mathbf{S}_l$, $\mathbf{\hat{B}}$, and $\mathbf{S}_h\!\cdot\!\mathbf{S}_l$  in the j-j basis.

\paragraph*{3. The D-wave:} For $L=2$ (D-wave), a similar analysis can be carried out. In the heavy–light meson case $S_h = S_l = \tfrac{1}{2}$, so again $S=0$ or $1$. Coupling $S$ with $L=2$ yields possible $J$ values $1,2,3$. The $L$–$S$ basis states can be labeled as $|{}^1D_2\rangle$ ($S=0$, $J=2$) and $|{}^3D_J\rangle$ ($S=1$, $J=1,2,3$). Thus $J=2$ occurs for both $S=0$ and $1$ (two basis states $|{}^1D_2\rangle$, $|{}^3D_2\rangle$ that can mix), while $J=1$ ($|{}^3D_1\rangle$) and $J=3$ ($|{}^3D_3\rangle$) are pure triplet states with no partner.

We begin by using Eq. [\ref{eq:couple-LS-J} ] to build the $L-S$ basis, and the results are

\begin{equation} 
|^{3}D_{1};1\rangle = \sqrt{\frac{3}{5}}|-\frac{1}{2},-\frac{1}{2},2\rangle - \frac{1}{2}\sqrt{\frac{3}{5}}	|\frac{1}{2},-\frac{1}{2},1\rangle- \frac{1}{2}	\sqrt{\frac{3}{5}}|-\frac{1}{2},\frac{1}{2},1\rangle + \frac{1}{\sqrt{10}}		|\frac{1}{2},\frac{1}{2},0\rangle.
\end{equation}

\begin{equation} 
|^{1}D_{2};2\rangle = \frac{1}{\sqrt{2}}|\frac{1}{2},-\frac{1}{2},2\rangle - \frac{1}{\sqrt{2}}	|-\frac{1}{2},\frac{1}{2},2\rangle
\end{equation}

\begin{equation} 
|^{3}D_{2};2\rangle = -\frac{1}{\sqrt{3}}|\frac{1}{2},-\frac{1}{2},2\rangle - \frac{1}{\sqrt{3}}	|-\frac{1}{2},\frac{1}{2},2\rangle+ 
\frac{1}{\sqrt{3}}	|\frac{1}{2},\frac{1}{2},1\rangle.
\end{equation}

\begin{equation} 
|^{3}D_{3};3\rangle = |\frac{1}{2},\frac{1}{2},2\rangle 
\end{equation}

Using these $L$–$S$ coupling basis, we compute the expectation values of the operators given in Eqs.~[\ref{eq:LS-operator}], [\ref{eq:B-tensor}], and [\ref{eq:spinspin-expect}] within the [$^{2}P_{J}$, $^{4}P_{J}$] basis for all allowed values of $J$, as listed below:\\

Subsequently, we compute the expectation values of the mass-splitting operators within the  {$L$–$S$}. These values, for specific values of $J$, are presented in the list below:

For J=1,
\begin{equation}
\text{$\langle\mathbf{L}.\mathbf{S}_l\rangle$=}-\frac{3}{2},\ \  
\text{$\langle\mathbf{L}.\mathbf{S}_h\rangle$=}-\frac{3}{2},\ \   
\text{$\langle\mathbf{\hat{B}}\rangle$=}-\frac{1}{2},\ \ 
\text{$\langle\mathbf{S}_l.\mathbf{S}_h\rangle$=}\frac{1}{4}.
\end{equation}	
For J=2,
\begin{equation}
\text{$\langle\mathbf{L}.\mathbf{S}_l\rangle$=}\left[
\begin{array}{cc}
0 & -\sqrt{\frac{3}{2}} \\
-\sqrt{\frac{3}{2}} & -\frac{1}{2} \\
\end{array}
\right],\ \   	
\text{$\langle\mathbf{L}.\mathbf{S}_h\rangle$=}\left[
\begin{array}{cc}
0 & \sqrt{\frac{3}{2}} \\
\sqrt{\frac{3}{2}} & -\frac{1}{2} \\
\end{array}
\right],\ \   
\text{$\langle\mathbf{\hat{B}}\rangle$=}\left[
\begin{array}{cc}
0 & 0 \\
0 & \frac{1}{2} \\
\end{array}
\right],\ \ 
\text{$\langle\mathbf{S}_l.\mathbf{S}_h\rangle$=}\left[
\begin{array}{cc}
-\frac{3}{4} & 0 \\
0 & \frac{1}{4} \\
\end{array}
\right].
\end{equation}	
For J=3,
\begin{equation}
\text{$\langle\mathbf{L}.\mathbf{S}_l\rangle$=}1,\ \  
\text{$\langle\mathbf{L}.\mathbf{S}_h\rangle$=}1,\ \   
\text{$\langle\mathbf{\hat{B}}\rangle$=}-\frac{1}{7},\ \ 
\text{$\langle\mathbf{S}_l.\mathbf{S}_h\rangle$=}\frac{1}{4}.
\end{equation}	

Following the same steps as the P-wave, we first diagonalize the dominant light-quark spin-orbit term $\mathbf{L}\!\cdot\!\mathbf{S}_l$. With $L=2$ and $S_l=\frac{1}{2}$, the total light quark angular momentum $j = L+S_l$ can be $j = \frac{3}{2}$ or $\frac{5}{2}$. The eigenvalues of $\mathbf{L}\!\cdot\!\mathbf{S}_l$ are $\lambda = -\tfrac{3}{2}$ for $j=\tfrac{3}{2}$ and $\lambda = +1$ for $j=\tfrac{5}{2}$. In the $J=2$ subspace, one finds the corresponding eigenvectors as mixtures of $|{}^1D_2\rangle$ and $|{}^3D_2\rangle$:
\begin{equation}
\lambda= -\frac{3}{2} :
|J=2, j=\frac{3}{2} \rangle =
\sqrt{\frac{2}{5}}  |^{1}D_{2}   \rangle 
+\sqrt{\frac{3}{5}} |^{3}D_{2}   \rangle,
\end{equation}

\begin{equation}
\lambda= 1 :
|J=2, j=\frac{5}{2} \rangle =
-\sqrt{\frac{3}{5}}  |^{1}D_{2}   \rangle 
+\sqrt{\frac{2}{5}} |^{3}D_{2}   \rangle,
\end{equation}
For $J=1$ and $J=3$, only the $|{}^3D_1\rangle$ and $|{}^3D_3\rangle$ states (with $j=\tfrac{3}{2}$ and $j=\tfrac{5}{2}$ respectively) exist, which are already eigenstates of $\mathbf{L}\!\cdot\!\mathbf{S}_l$ with $\lambda = -\tfrac{3}{2}$ for $J=1$ and $\lambda = +1$ for $J=3$. Hence, the following relations hold:

\begin{equation}
|J=1, j=\frac{3}{2} \rangle =  |^{3}D_{1}\rangle,
\end{equation}

\begin{equation}
|J=3, j=\frac{5}{2} \rangle =  |^{3}D_{3}\rangle,
\end{equation}

Using these $|j, J\rangle$ basis, the expectation values of the operators in the D-wave  can be computed in a similar manner. These computed values then contribute to the spin-dependent splitting of the heavy–light meson spectra of D-wave.

\subsection*{B. Doubly Heavy Baryons}
\paragraph*{1. The S-wave:} We now turn to doubly heavy baryons containing a heavy $QQ'$ diquark and a light quark. We take the diquark to be in the spin-1 (vector) configuration, $S_h=1$, as appropriate for a symmetric $QQ'$ state in a color $\bar{\mathbf{3}}$ (antisymmetric) combination. The light quark has $S_l=\tfrac{1}{2}$. For the ground-state baryons with $L=0$, only the spin-spin contact interaction between the heavy diquark and light quark is relevant. Analogous to Eq.~\eqref{eq:spinspin-expect}, one finds 
\[ \langle\mathbf{S}_h\!\cdot\!\mathbf{S}_l\rangle \;=\; \frac{1}{2}\Big[\,S(S+1) - S_h(S_h+1) - S_l(S_l+1)\,\Big], \] 
with $S=|S_h - S_l|,\,\dots,\,S_h+S_l$. For $S_h=1$ and $S_l=\frac{1}{2}$, the possible total spin of the baryon’s two-body subsystem is $S=\tfrac{1}{2}$ or $\tfrac{3}{2}$. In the $L=0$ ground state, the light quark and diquark are predominantly in the relative $S$-wave, so the total $J^P$ of the baryon is $J^P = S^+_{\!}$. The spin–spin operator $\langle\mathbf{S}_h\!\cdot\!\mathbf{S}_l\rangle$ yields $-1$ for the $S=\tfrac{1}{2}$ configuration and $+\tfrac{1}{2}$ for $S=\tfrac{3}{2}$, indicating that the heavier-spin state is elevated. 

\paragraph*{2. The P-wave:} For $L=1$ orbital excitation of a doubly heavy baryon, we have three angular momenta: the orbital $L=1$, the heavy diquark spin $S_h=1$, and the light quark spin $S_l=\frac{1}{2}$. In the $L$–$S$ coupling scheme, $S_h$ and $S_l$ first combine to total spin $S$, which can be $\frac{1}{2}$ or $\frac{3}{2}$. Coupling $S$ with $L=1$ yields the baryon total $J$. The $L$–$S$ basis states can be denoted $|{}^2P_J\rangle$ for $S=\tfrac{1}{2}$ and $|{}^4P_J\rangle$ for $S=\tfrac{3}{2}$, with $J$ taking the values $\frac{1}{2}, \frac{3}{2},  \frac{5}{2}$. They can be constructed in terms of $|S_{h_3},\,S_{l_3},\,L_3\rangle$ using Eq. [\ref{eq:couple-LS-J} ] as 

\begin{equation} 
|^{2}P_{1/2};1/2\rangle = \frac{\sqrt{2}}{3}|-\frac{1}{2},0,1\rangle -\frac{\sqrt{2}}{3}|-\frac{1}{2},1,0\rangle-\frac{2}{3}|\frac{1}{2},-1,1\rangle+\frac{1}{3}|\frac{1}{2},0,0\rangle
\end{equation}

\begin{equation} 
|^{4}P_{1/2};1/2\rangle = \frac{1}{3}|-\frac{1}{2},0,1\rangle -\frac{1}{3}|-\frac{1}{2},	1,0\rangle
+\frac{1}{3\sqrt{2}}|\frac{1}{2},-1,1\rangle
-\frac{\sqrt{2}}{3}|\frac{1}{2},0,0\rangle
+\frac{1}{\sqrt{2}}|\frac{1}{2},1,-1\rangle
\end{equation}

\begin{equation} 
|^{2}P_{3/2};3/2\rangle = -\sqrt{\frac{2}{3}}|-\frac{1}{2},1,1\rangle +\frac{1}{\sqrt{3}}|\frac{1}{2},0,1\rangle
\end{equation}

\begin{equation} 
|^{4}P_{3/2};3/2\rangle = -\sqrt{\frac{2}{15}}|-\frac{1}{2},1,1\rangle -\frac{2}{\sqrt{15}}|\frac{1}{2},0,1\rangle
+\sqrt{\frac{3}{5}}|\frac{1}{2},1,0\rangle
\end{equation}

\begin{equation} 
|^{4}P_{5/2};5/2\rangle = |\frac{1}{2},1,1\rangle 
\end{equation}
In particular, there are two $J=\tfrac{1}{2}$ states ($|{}^2P_{\frac{1}{2}}\rangle, |{}^4P_{\frac{1}{2}}\rangle$) and two $J=\tfrac{3}{2}$ states ($|{}^2P_{\frac{3}{2}}\rangle, |{}^4P_{\frac{3}{2}}\rangle$) that can mix , while $J=\tfrac{5}{2}$ has only the $|{}^4P_{\frac{5}{2}}\rangle$ state (with $S=\frac{3}{2}$).

The expression for spin-dependent operators (Eqs.~[\ref{eq:LS-operator}], [\ref{eq:B-tensor}], and [\ref{eq:spinspin-expect}] )are analogous to the mesonic case, with only difference that the  $\mathbf{S}_h$ now represents the spin of heavy diquark  and the $\mathbf{S}_l$ represents the spin of light quark.

Using these expressions, we can calculate the  expectation value for the spin-dependent operators within the [$^{2}P_{J}$, $^{4}P_{J}$] basis for all allowed values of $J$, as listed below:\\
For J=1/2,
\begin{equation}
\text{$\langle\mathbf{L}.\mathbf{S}_l\rangle$=}\left[
\begin{array}{cc}
\frac{1}{3} & -\frac{\sqrt{2}}{3} \\
-\frac{\sqrt{2}}{3} & -\frac{5}{6} \\
\end{array}
\right],\ \  
\text{$\langle\mathbf{L}.\mathbf{S}_h\rangle$=}\left[
\begin{array}{cc}
-\frac{4}{3} & \frac{\sqrt{2}}{3} \\
\frac{\sqrt{2}}{3} & -\frac{5}{3} \\
\end{array}
\right],\ \   
\text{$\langle\mathbf{\hat{B}}\rangle$=}\left[
\begin{array}{cc}
0 & -\frac{1}{\sqrt{2}} \\
-\frac{1}{\sqrt{2}} & -1 \\
\end{array}
\right],\ \ 
\text{$\langle\mathbf{S}_l.\mathbf{S}_h\rangle$=}\left[
\begin{array}{cc}
-1 & 0 \\
0 & \frac{1}{2} \\
\end{array}
\right].
\end{equation}	
For J=3/2,
\begin{equation}
\text{$\langle\mathbf{L}.\mathbf{S}_l\rangle$=}\left[
\begin{array}{cc}
-\frac{1}{6} & -\frac{\sqrt{5}}{3} \\
-\frac{\sqrt{5}}{3} & -\frac{1}{3} \\
\end{array}
\right],\ \  
\text{$\langle\mathbf{L}.\mathbf{S}_h\rangle$=}\left[
\begin{array}{cc}
\frac{2}{3} & \frac{\sqrt{5}}{3} \\
\frac{\sqrt{5}}{3} & -\frac{2}{3} \\
\end{array}
\right],\ \   
\text{$\langle\mathbf{\hat{B}}\rangle$=}\left[
\begin{array}{cc}
0 & \frac{1}{2 \sqrt{5}} \\
\frac{1}{2 \sqrt{5}} & \frac{4}{5} \\
\end{array}
\right],\ \ 
\text{$\langle\mathbf{S}_l.\mathbf{S}_h\rangle$=}\left[
\begin{array}{cc}
-1 & 0 \\
0 & \frac{1}{2} \\
\end{array}
\right].
\end{equation}	
For J=5/2,
\begin{equation}
\text{$\langle\mathbf{L}.\mathbf{S}_l\rangle$=}\frac{1}{2},\ \  
\text{$\langle\mathbf{L}.\mathbf{S}_h\rangle$=}1,\ \   
\text{$\langle\mathbf{\hat{B}}\rangle$=}-\frac{1}{5},\ \ 
\text{$\langle\mathbf{S}_l.\mathbf{S}_h\rangle$=}\frac{1}{2}.
\end{equation}	
The heavy $QQ'$ diquark, which is much more massive, is analogous to a heavy antiquark in a meson. Thus, in the heavy-diquark limit ($M_{QQ} \gg m_q$), the $\mathbf{L}\!\cdot\!\mathbf{S}_l$ term associated with the light quark dominates the spin-orbit splitting. We therefore employ a $J$–$j$ coupling scheme where we first couple $\mathbf{L}$ with $\mathbf{S}_l$ to form $j = L + S_l$. For $L=1$ and $S_l=\frac{1}{2}$, one gets $j=\frac{1}{2}$ or $\frac{3}{2}$. The baryon states can then be labeled $|J,\,j\rangle$, and one can treat the remaining $\mathbf{L}\!\cdot\!\mathbf{S}_h$ and tensor interactions (involving the heavy diquark’s spin) as perturbations. Diagonalizing $\mathbf{L}\!\cdot\!\mathbf{S}_l$ within a given $J$ subspace yields eigenvalues $\lambda=-1$ for $j=\frac{1}{2}$ and $\lambda=+\tfrac{1}{2}$ for $j=\frac{3}{2}$ . For example, the two physical $J=\frac{1}{2}$ states can be expressed as 

\begin{equation}
\lambda= -1 :
|J=\frac{1}{2}, j=1/2 \rangle =
\frac{1}{3}  |^{2}P_{1/2}   \rangle 
+\frac{2\sqrt{2}}{3}  |^{4}P_{1/2}   \rangle,
\end{equation}
\begin{equation}
\lambda= \frac{1}{2} :
|J=\frac{1}{2}, j=3/2 \rangle =
-\frac{2\sqrt{2}}{3}|^{2}P_{1/2}   \rangle 
+\frac{1}{3}  |^{4}P_{1/2}   \rangle,
\end{equation}
in terms of the $L$–$S$ basis $\{|{}^2P_{1/2}\rangle,\,|{}^4P_{1/2}\rangle\}$. These correspond to $j=\frac{1}{2}$ and $j=\frac{3}{2}$ eigenstates of $\mathbf{L}\!\cdot\!\mathbf{S}_l$, with eigenvalues $-1$ and $+\tfrac{1}{2}$ respectively.

Likewise, for the $J=\frac{3}{2}$ sector we get 
\begin{equation}
\lambda= -1 :
|J=\frac{3}{2}, j=1/2 \rangle =
\frac{2}{3}  |^{2}P_{3/2}   \rangle 
+\frac{\sqrt{5}}{3}  |^{4}P_{3/2}   \rangle,
\end{equation}
\begin{equation}
\lambda= \frac{1}{2} :
|J=\frac{3}{2}, j=3/2 \rangle =
-\frac{\sqrt{5}}{3} |^{2}P_{3/2}   \rangle 
+\frac{2}{3}  |^{4}P_{3/2}   \rangle,
\end{equation}

While the $J=\frac{5}{2}$ state remains purely $|{}^4P_{5/2}\rangle$ with $j=\frac{3}{2}$
\begin{equation}
|J=\frac{5}{2}, j=3/2 \rangle =
|^{4}P_{5/2}\rangle 
\end{equation}

Using these states, we determine the expectation values of the operators in the $J, j \rangle$ basis and list the results in Table \ref{table2}, which can be used to calculate the P-wave mass splittings.

\paragraph*{3. The D-wave:} Finally, we consider the $L=2$ (D-wave) excitations of doubly heavy baryons. The heavy diquark has $S_h=1$ and the light quark $S_l=\frac{1}{2}$ as before. In the $L$–$S$ scheme, these combine to $S=\frac{1}{2}$ or $\frac{3}{2}$, which then couple with $L=2$ to yield possible total $J$ values $\frac{1}{2}, \frac{3}{2}, \frac{5}{2}, \frac{7}{2}$. The corresponding $L-S$ basis for $D$-wave states can be formed using Eq. [\ref{eq:couple-LS-J}] as

\begin{equation} 
|^{4}D_{1/2};1/2\rangle = 
-\sqrt{\frac{2}{5}}|-\frac{1}{2},-1,2\rangle 
+\frac{1}{\sqrt{5}}|-\frac{1}{2},0,1\rangle 
-\frac{1}{\sqrt{15}}|-\frac{1}{2},1,0\rangle 
+\frac{1}{\sqrt{10}}|\frac{1}{2},-1,1\rangle 
-\sqrt{\frac{2}{15}}|\frac{1}{2},0,0\rangle 
+\frac{1}{\sqrt{10}}|\frac{1}{2},1,-1\rangle 
\end{equation}

\begin{equation} 
|^{2}D_{3/2};3/2\rangle = 
\frac{2}{\sqrt{15}}|-\frac{1}{2},0,2\rangle 
-\sqrt{\frac{2}{15}}|-\frac{1}{2},1,1\rangle 
-2\sqrt{\frac{2}{15}}|\frac{1}{2},-1,2\rangle 
+\frac{1}{\sqrt{15}}|\frac{1}{2},0,1\rangle 
\end{equation}

\begin{equation} 
|^{4}D_{3/2};3/2\rangle = 
\frac{2}{\sqrt{15}}|-\frac{1}{2},0,2\rangle 
-\sqrt{\frac{2}{15}}|-\frac{1}{2},1,1\rangle 
+\sqrt{\frac{2}{15}}|\frac{1}{2},-1,2\rangle 
-\frac{2}{\sqrt{15}}|\frac{1}{2},0,1\rangle 
+\frac{1}{\sqrt{5}}|\frac{1}{2},1,0\rangle 
\end{equation}

\begin{equation} 
|^{2}D_{5/2};5/2\rangle = -\sqrt{\frac{2}{3}}|-\frac{1}{2},1,2\rangle +\frac{1}{\sqrt{3}}|\frac{1}{2},0,2\rangle 
\end{equation}

\begin{equation} 
|^{4}D_{5/2};5/2\rangle = 
-\frac{2}{\sqrt{21}}|-\frac{1}{2},1,2\rangle 
-2\sqrt{\frac{2}{21}}|\frac{1}{2},0,2\rangle 
+\sqrt{\frac{3}{7}}|\frac{1}{2},1,1\rangle 
\end{equation}

\begin{equation} 
|^{4}D_{7/2};7/2\rangle = 
|\frac{1}{2},1,2\rangle
\end{equation}

Then, the expectation of the mass-splitting operators given in Eqs.~[\ref{eq:LS-operator}], [\ref{eq:B-tensor}], and [\ref{eq:spinspin-expect}], in basis [$|{}^2D_J\rangle$, $|{}^4D_J\rangle$], are evaluated, and the results are as follows:

For J=1/2,
\begin{equation}
\text{$\langle\mathbf{L}.\mathbf{S}_l\rangle$=}-\frac{3}{2},\ \  
\text{$\langle\mathbf{L}.\mathbf{S}_h\rangle$=}-3,\ \   
\text{$\langle\mathbf{\hat{B}}\rangle$=}-1,\ \ 
\text{$\langle\mathbf{S}_l.\mathbf{S}_h\rangle$=}\frac{1}{2}.
\end{equation}

For J=3/2,
\begin{equation}
\text{$\langle\mathbf{L}.\mathbf{S}_l\rangle$=}\left[
\begin{array}{cc}
\frac{1}{2} & -1 \\
-1 & -1 \\
\end{array}
\right],\ \  
\text{$\langle\mathbf{L}.\mathbf{S}_h\rangle$=}\left[
\begin{array}{cc}
-2 & 1 \\
1 & -2 \\
\end{array}
\right],\ \   
\text{$\langle\mathbf{\hat{B}}\rangle$=}\left[
\begin{array}{cc}
0 & -\frac{1}{2} \\
-\frac{1}{2} & 0 \\
\end{array}
\right],\ \ 
\text{$\langle\mathbf{S}_l.\mathbf{S}_h\rangle$=}\left[
\begin{array}{cc}
-1 & 0 \\
0 & \frac{1}{2} \\
\end{array}
\right].
\end{equation}
For J=5/2,
\begin{equation}
\text{$\langle\mathbf{L}.\mathbf{S}_l\rangle$=}\left[
\begin{array}{cc}
-\frac{1}{3} & -\frac{\sqrt{14}}{3} \\
-\frac{\sqrt{14}}{3} & -\frac{1}{6} \\
\end{array}
\right],\ \  
\text{$\langle\mathbf{L}.\mathbf{S}_h\rangle$=}\left[
\begin{array}{cc}
\frac{4}{3} & \frac{\sqrt{14}}{3} \\
\frac{\sqrt{14}}{3} & -\frac{1}{3} \\
\end{array}
\right],\ \   
\text{$\langle\mathbf{\hat{B}}\rangle$=}\left[
\begin{array}{cc}
0 & \frac{1}{\sqrt{14}} \\
\frac{1}{\sqrt{14}} & \frac{5}{7} \\
\end{array}
\right],\ \ 
\text{$\langle\mathbf{S}_l.\mathbf{S}_h\rangle$=}\left[
\begin{array}{cc}
-1 & 0 \\
0 & \frac{1}{2} \\
\end{array}
\right].
\end{equation}	
For J=7/2,
\begin{equation}
\text{$\langle\mathbf{L}.\mathbf{S}_l\rangle$=}1,\ \  
\text{$\langle\mathbf{L}.\mathbf{S}_h\rangle$=}2,\ \   
\text{$\langle\mathbf{\hat{B}}\rangle$=}-\frac{2}{7},\ \ 
\text{$\langle\mathbf{S}_l.\mathbf{S}_h\rangle$=}\frac{1}{2}.
\end{equation}

As before, we now switch to the $J$–$j$ coupling by diagonalizing $\mathbf{L}\!\cdot\!\mathbf{S}_l$ matrix.  This yields $j=\frac{3}{2}$ (with $\lambda=-\tfrac{3}{2}$) and $j=\frac{5}{2}$ (with $\lambda=+1$) as the light-quark total angular momenta. Consequently, the two $J=\frac{3}{2}$ basis states ($|{}^2D_{3/2}\rangle$, $|{}^4D_{3/2}\rangle$) mix to form $|J=\frac{3}{2},\,j=\frac{3}{2}\rangle$ and $|J=\frac{3}{2},\,j=\frac{5}{2}\rangle$ eigenstates of $\mathbf{L}\!\cdot\!\mathbf{S}_l$

\begin{equation}
\lambda= -\frac{3}{2} :
|J=\frac{3}{2}, j=3/2 \rangle =
\frac{1}{\sqrt{5}}|^{2}D_{3/2}   \rangle 
+\frac{2}{\sqrt{5}}|^{4}D_{3/2}   \rangle,
\end{equation}

\begin{equation}
\lambda= 1 :
|J=\frac{3}{2}, j=5/2 \rangle =
-\frac{2}{\sqrt{5}}  |^{2}D_{3/2}   \rangle 
+\frac{1}{\sqrt{5}}  |^{4}D_{3/2}   \rangle,
\end{equation}

Similarly the $J=\frac{5}{2}$ basis ($|{}^2D_{5/2}\rangle$, $|{}^4D_{5/2}\rangle$) mix to form $|J=\frac{5}{2},\,j=\frac{3}{2}\rangle$ and $|J=\frac{5}{2},\,j=\frac{5}{2}\rangle$ eigenstates of $\mathbf{L}\!\cdot\!\mathbf{S}_l$ as below:

\begin{equation}
\lambda= -\frac{3}{2} :
|J=\frac{5}{2}, j=3/2 \rangle =
2\sqrt{\frac{2}{15}}|^{2}D_{5/2}   \rangle 
+\sqrt{\frac{7}{15}}|^{4}D_{5/2}   \rangle,
\end{equation}

\begin{equation}
\lambda= 1 :
|J=\frac{5}{2}, j=5/2 \rangle =
-\sqrt{\frac{7}{15}}  |^{2}D_{5/2}   \rangle 
+2\sqrt{\frac{2}{15}}  |^{4}D_{5/2}   \rangle,
\end{equation}

The extreme $J$ values, $J=\frac{1}{2}$ and $J=\frac{7}{2}$, correspond uniquely to $|{}^4D_{1/2}\rangle$ and $|{}^4D_{7/2}\rangle$ 
\begin{equation}
|J=\frac{1}{2}, j=3/2 \rangle =
|^{4}D_{1/2}   \rangle ,
\end{equation}
\begin{equation}
|J=\frac{7}{2}, j=5/2 \rangle =
|^{4}D_{7/2}   \rangle ,
\end{equation}

Ultimately, we calculate the expectation values of the mass-splitting operators within the $|J, j\rangle$ basis and compile the findings in a table \ref{table2} for detailed analysis and reference.

\end{widetext}
	\bibliography{apssamp}
\end{document}